\documentclass[journal=jpcafh,manuscript=article]{achemso}
\usepackage{chemformula} 
\usepackage[utf8]{inputenc}
\usepackage[english]{babel}
\usepackage[T1]{fontenc}
\usepackage{amsmath}
\usepackage{amsfonts}
\usepackage{amssymb}
\usepackage{lmodern}
\usepackage{float}
\usepackage{graphicx}
\usepackage{xcolor}
\usepackage{siunitx}
\usepackage{subcaption}
\title{Hydrogenation of C$_{24}$ carbon clusters : structural diversity and energetic properties}

\newcommand{\cbn}[1]{C$_{#1}$}
\newcommand{\hbn}[1]{H$_{#1}$}
\newcommand{\cbnhm}[2]{C$_{#1}$H$_{#2}$}

\newcommand{\angstrom}{\text{\normalfont\AA}}

\SectionNumbersOn

\author{Paula Pla}
\affiliation{Departamento de Qu\'imica, Universidad Aut\'onoma de Madrid, M\'odulo 13, 28049 Madrid, Spain}
\author{Cl\'ement Dubosq}
\affiliation{Laboratoire de Chimie et Physique Quantiques (LCPQ), F\'ed\'eration FERMI, Univ. Toulouse UT3 \& CNRS, UMR5626, 118 Route Narbonne, F-31062 Toulouse, France}
\author{Mathias Rapacioli}
\affiliation{Laboratoire de Chimie et Physique Quantiques (LCPQ), F\'ed\'eration FERMI, Univ. Toulouse UT3 \& CNRS, UMR5626, 118 Route Narbonne, F-31062 Toulouse, France}
\author{Evgeny Posenitskiy}
\affiliation{Laboratoire Collisions Agr\'egats et R\'eactivit\'e (LCAR), Universit\'e de Toulouse (UPS) and CNRS, UMR5589, 118 Route de Narbonne, F-31062 Toulouse, France}
\author{Manuel Alcam\'i}
\affiliation{Departamento de Qu\'imica, Universidad Aut\'onoma de Madrid, M\'odulo 13, 28049 Madrid, Spain}
\alsoaffiliation{Instituto Madrile\~no de Estudios Avanzados en Nanociencia (IMDEA-Nanociencia), 28049 Madrid, Spain}
\alsoaffiliation{Institute for Advanced Research in Chemical Sciences (IAdChem), Universidad Aut\'onoma de Madrid, 28049 Madrid, Spain}
\author{Aude Simon}
\affiliation{Laboratoire de Chimie et Physique Quantiques (LCPQ), F\'ed\'eration FERMI, Univ. Toulouse UT3 \& CNRS, UMR5626, 118 Route Narbonne, F-31062 Toulouse, France}
\email{aude.simon@irsamc.ups-tlse.fr}
\phone{+123 (0)123 4445556}
\fax{+123 (0)123 4445557}

\title{Hydrogenation of C$_{24}$ carbon clusters : structural diversity and energetic properties}
\abbreviations{IR,NMR,UV}

\begin{document}
\begin{tocentry}


\includegraphics[height=3.5cm]{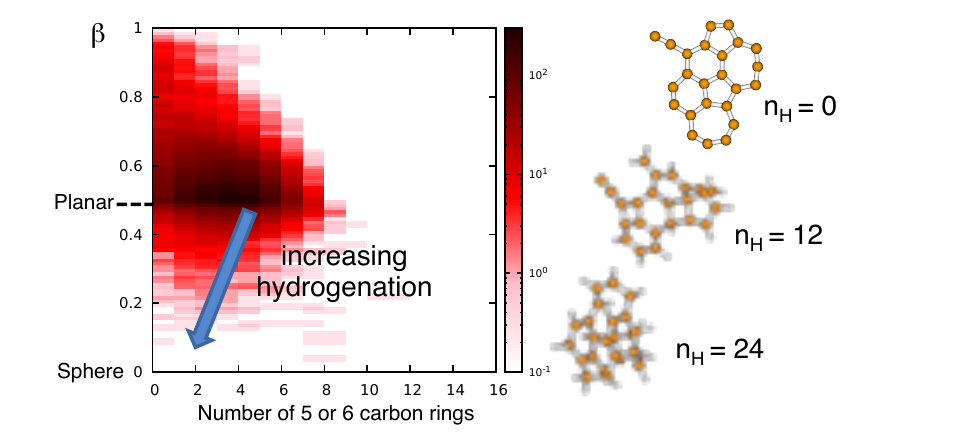}




\end{tocentry}



\begin{abstract}
  This work aims at exploring the potential energy surfaces of \cbnhm{24}{n=0,6,12,18,24} up to 20-25\,eV using the genetic algorithm in combination with the density functional based tight binding (DFTB) potential. The structural diversity of the non fragmented structures was analysed using order parameters which were chosen as the number of 5 or 6 member rings
  and the asphericity constant $\beta$.
  The most abundant and lowest energy population was found to correspond to a  flakes population, constituted of isomers of  variable shapes 
  possessing a large number of 5 or 6-carbon rings.  
This population is characterized by a larger number of spherical isomers when $n_H/n_C$ increases.  Simultaneously, the fraction of the pretzels population constituted of spherical isomers possessing fewer 5 or 6 carbon ring cycles increases. 
For all hydrogenation rates, the fraction of cages population, possessing the largest number of 5 or 6-carbon rings remains extremely minor while the branched population, characterized by the smallest number of 5 or 6-carbon rings, is the highest energy population for all  $n_H/n_C$ ratios.   For all \cbnhm{24}{n=0,6,12,18,24} clusters, a detailed study of the evolution of the carbon ring size distribution as a function of energy clearly shows that the  stability is correlated to the number of 6-carbon rings. A similar study for hybridization $sp^n$ (n=1-3)  shows that the number of $sp^1$ carbon atoms increases with energy while globally the number of $sp^3$ carbon atoms increases with $n_H/n_C$. 
  The average values of the  ionization potentials of all populations, obtained at the self-consistent charge DFTB level, were found to decrease when  $n_H/n_C$ increases, ranging from 7.9\,eV down to 
  6.4\,eV. We correlated this trend to geometric and electronic factors, in particular to carbon atoms hybridization $sp^n$ (n=1-3).  
These results are of astrophysical interest as they should be taken into account in astrophysical models especially regarding the role of carbonaceous species in  the gas ionization. 

\end{abstract}

\section{Introduction}

Among cosmic elements, 
carbon is the second most abundant one (after oxygen) among those able to form bonds of various natures and thus to participate in cosmic molecular complexity through top-down \cite{Jones2016} (destruction) and bottom-up  \cite{Joblin2020} (growth) processes 
Given its rich allotropy, a diversity of interstellar carbonaceous molecules and dust grains were shown to contain a large fraction of the carbon budget in the galaxy.   We can cite nano-diamonds, fullerenes, polyaromatic structures or  (hydrogenated)- amorphous carbon structures \cite{Dartois2019-rev}.  These species were identified thanks to the interplay between spectroscopic observations, experimental and theoretical studies. However, more studies are mandatory to refine the assignment of  interstellar features attributed to carbon compounds but whose precise structures remain elusive. \\
First, the need to identify the carriers of the aromatic infrared bands (AIBs), a set of mid-IR emission bands observed ubiquitously in the interstellar medium (ISM) \cite{leger84,aib2} motivated  many experimental and theoretical studies.  In particular, quantum chemistry revealed a powerful tool to compute many IR spectra of individual  polyclic aromatic hydrocarbons (PAHs) of different sizes and shapes, neutral and charged, incorporating heteroatoms or complexed to a metal atom
 \cite{2009Bausch,2010Simon}. 
However, the AIBs are constituted of bands but also of broad plateaus. In particular, IR spectroscopic observations of fullerene-rich planetary nebulae (PNe) exhibit a
 broad plateau with substructure in the 6--9 \si{\micro\meter} \cite{Hernandez2010,Hernandez2011,Hernandez2011b,Hernandez2012,Salas_ApJ2012}. The origin of this plateau remains elusive and various hypotheses were formulated about its possible carriers, including hydrogenated amorphous carbon (HAC) compounds \cite{Salas_ApJ2012}, very small grains, possibly PAH clusters \citep{tielensrev,Buss2013,Rapacioli2005}. Similarly, the  8 and 12 \si{\micro\meter} plateau features of some proto-planetary nebulae (PPNe) were assigned to the vibrations of alkanes or alkyle side groups pertaining to large carbonaceous particles \cite{Kwok2001,Hrivnak2000}, whose details also remain unclear.  Very recently, quantum calculations showed that the isomers' population of C$_{60}$ containing the most spherical isomers with a high fraction of sp$^2$ carbon atoms was a good candidate to carry the 6--9 \si{\micro\meter} plateau \cite{dubosq_mapping_2019-1} whereas  
8 and 12 \si{\micro\meter} plateau could hardly be reproduced. These observations and their various interpretations therefore motivate further studies to unravel the molecular nature of the carriers of these spectral features, in particular addressing the possible contribution of hydrogenated carbon compounds. \\
%
Second, the UV bump \cite{stecher1965interstellar} is a broad ultraviolet absorption bump observed on the ISM extinction curves in systems ranging from the Milky Way to high-redshift galaxies, with a stable position centered at 217.5 nm, although showing variations in intensity and width 
in different lines of sight 
  [see Ref. \cite{2019Fitzpatrick} and references therein].

Interestingly, the variations in the UV bump width and shape have been related to the presence of defects within the sp$^2$ conjugated carbon network. Several theoretical and experimental studies were subsequently carried out to unravel the structural origin of these variations. The effects of disorder on the optical absorption spectra of graphitic grains were modeled  \cite{papoular2013effects},  introducing disorder by replacing sp$^2$ carbons by sp$^3$ and sp$^1$ atoms. Interestingly, these authors highlighted an effect 
on the width of the computed band that they located at 5.77\,eV, in agreement with Draine's model \citep{draine1984optical}.
More generally, the contribution of more disordered species was further elaborated under the form of hydrogenated amorphous  carbon (HAC) \cite{jones1990structure,gadallah2011uv} and soot particles [see Ref. \cite{gavilan2017polyaromatic} and references therein]. In an attempt to establish correlations between carbon grain morphology and spectral data,  Rotundi {\it et al.} \cite{Rotundi1998} proposed that carbon nanostructures ordered on the micrometer scale could be better candidates to interpret the UV bump rather than graphitic particles. For instance, recent quantum modeling showed that  a population  of carbon spherical isomers with  a high fraction of conjugated sp$^2$ atoms could contribute  
to the astronomical UV bump  \cite{dubosq_quantum_2020}. \\
 
  In previous studies by Dubosq et al. \cite{dubosq_mapping_2019-1,dubosq_quantum_2020}, the diversity of pure carbon clusters was investigated as well as their IR and UV spectral properties. 
 However, as mentioned hereabove, some astronomical features could not be accounted for and the presence of hydrogenated carbon clusters must be considered. In this context, we now propose to explore and analyse the structural diversity of hydrogenated carbon clusters \cbnhm{24}{n=0,6,12,18,24} focusing on the lowest energy isomers. Their IR and UV visible features will be presented in future work. We  present the theoretical procedure in Section \ref{sec:theo}, its application to a test-case system, namely C$_{24}$, together with a discussion about appropriate order parameters in Section \ref{testcase}. Finally, the  results are reported in  Section \ref{sec:res}.\\


\section{Theoretical methods}\label{sec:theo}

\subsection{Electronic structure calculations}

We have used the density functional based tight binding (DFTB) \cite{dftb1,dftb2,elstner1998self} method to describe the electronic system. It results from  a compromise between accuracy and computational cost. This scheme  allows to describe carbonaceous systems with various hybridization orders and C-H chemical bonding, while preserving a computational cost much lower than {\it ab initio} methods.
 In the present work, we performed global exploration using the zeroth order DFTB (DFTB0) hamiltonian \cite{dftb1,dftb2} and subsequent local optimization using the second order (self consistent charge or SCC-) DFTB with the  mio set of  parameters \cite{elstner1998self},  an empirical dispersion correction contribution \cite{DFTB_CM3} and a Fermi temperature of 500 K to avoid some SCC convergence issues. All DFTB calculations were performed with the deMonNano code \cite{demonnano}.


\subsection{Genetic algorithm exploration}






The structures of \cbn{24}, \cbnhm{24}{6}, \cbnhm{24}{12}, \cbnhm{24}{18} and \cbnhm{24}{24} clusters obtained in this work were generated using a genetic algorithm (GA). 
GAs, which are based on the principles of evolution by natural selection,\cite{goldberg1989ga,forrest1993ga} are known to be
efficient methods for the search of the global minimum structure of clusters and nanoparticles.\cite{johnston2003ga}
In this work, the Atomic Simulation Environment (ASE) software package \cite{larsen2017ase} has been interfaced with the deMonNano code \cite{demonnano}.
The GA module of ASE has been used to obtain the minimum energy structures of different clusters such as metal clusters in MOFs,\cite{vilhelmsen2012asegamofs} supported nanostructures\cite{vilhelmsen2014aseganano} or mixed metal halide ammines.\cite{jensen2014asegaammines}. \\

The idea to explore the potential energy surface (PES) used herein was to benefit from the stochastic nature of genetic algorithms to create a population of structures of each cluster stoichiometry. 
 Structures are relaxed at the DFTB0 level during the run of the GA as stated below. The final structures obtained by the GA/DFTB0 algorithm are further locally optimized at the SCC-DFTB level. 

The details of a GA run are the following. First, we have to generate a set of random starting candidates (initial population) which was fixed at 20 structures. Each of these structures was generated by randomly positioning a set of carbon and hydrogen atoms in a box of 11.2 by 11.2 by 5 \angstrom{}. Second, the structures are relaxed using forces computed at the DFTB0 level and relaxation stops when gradient value is lower than 0.01 eV/\angstrom{} or when the number of steps reaches 1000.  This initial population constitutes  
the first parent population.
The GA run then consists in $n_c$ cycles. At each cycle, this population is updated by considering $n_o$ offsprings. To generate an offspring,
two structures are randomly selected from the parent population and each of them is cut into two fragments. Two fragments originating from two different parent structures are combined and relaxed into a new structure. 
In addition, three types of mutation can occur during the generation of the offspring candidate with a selected probability of 0.3. One of these mutations is a mirror mutation in which one of the fragments is mirrored. The second one is a rattle mutation in which 40$\%$ of the atoms in the structure are translated by a random distance below 0.8 \AA\,.
in a random direction. The third one which works only when different types of atoms compose the structure is permutation of the atomic number of one third of the atoms in the structure. By repeating these processes at each cycle, the population that will serve to produce the new generation of structures, is updated with the most stable structures. 
We keep track of all the DFTB0 relaxed geometries at each cycle to include them in the final population. 

In practice, we performed 200 independent GA runs using the DFTB0 potential for each hydrogenated cluster (50 for \cbn{24}) each of them consisting of $n_c=$36 cycles (30 for \cbn{24}) of $n_o=$ 30 new structures. 
 This number of cycles was set to be a good parameter to reach the global minimum structure but not to perform many more cycles once the minimum structure is reached. For each size, this leads to 200*36*30=216000 structures (45000 for $C_{24}$). 

Some of the structures obtained with the GA are fragmented structures which are removed from the final set. The MolMod python library was used to eliminate the fragmented structures based on their corresponding molecular graph.\cite{molmod} 
The non-fragmented structures were subsequently optimized at the SCC-DFTB level.
Finally, all isomers are ranked by energy and each structure differing by less than 1.0e$^{-6}$ au with respect to the previous one is considered as redundant and discarded.


\section{Benchmark of the GA/DFTB algorithm and definition of structural families}
\label{testcase}
Before generating sets of carbonaceous clusters for different H/C ratios, we performed preliminary simulations to check the efficiency of the GA in terms of PES exploration and thus diversity of structures. We chose \cbn{24} as its PES exploration followed by structural and spectral analysis was achieved in our previous works \cite{dubosq_mapping_2019-1,dubosq_quantum_2020}. In those studies, the PES exploration was performed 
using parallel tempering molecular dynamics (PTMD) \cite{Sugita1999} with the REBO force field \cite{Bonnin_2019}, 
followed by local SCC-DFTB optimization. 
Despite the fact that, in both studies [Ref. \cite{dubosq_mapping_2019-1} and the present work], 
 the final optimizations were conducted  at the same level of theory, they are expected to present differences arising from the two global exploration schemes (PTMD-REBO vs GA-DFTB0).

\begin{figure}[htbp!]
          (i) GA/DFTB structures\\
        \includegraphics[width=0.8\linewidth]{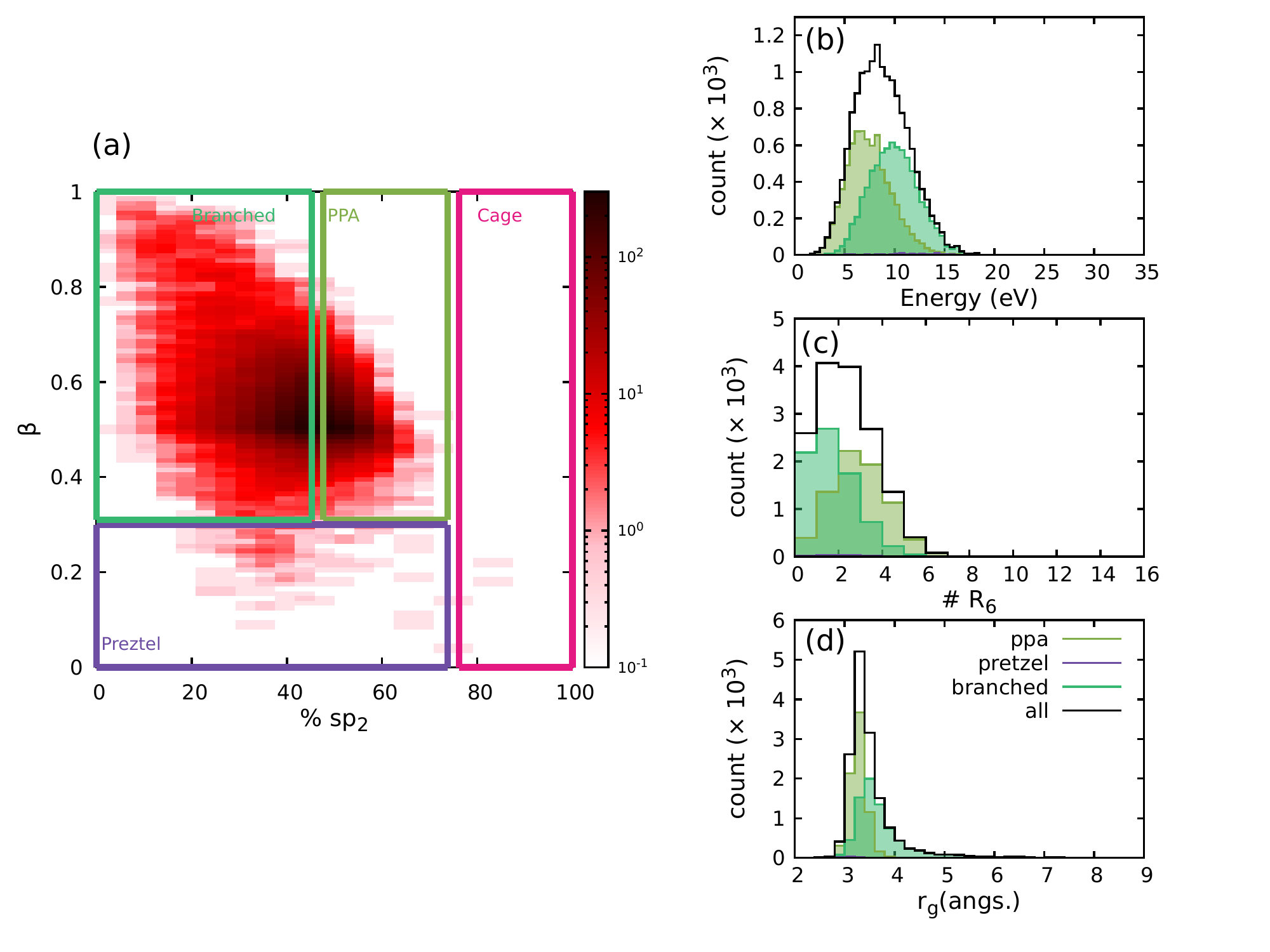} \\
        (ii)  PTMD/REBO structures \\
         \includegraphics[width=0.8\linewidth]{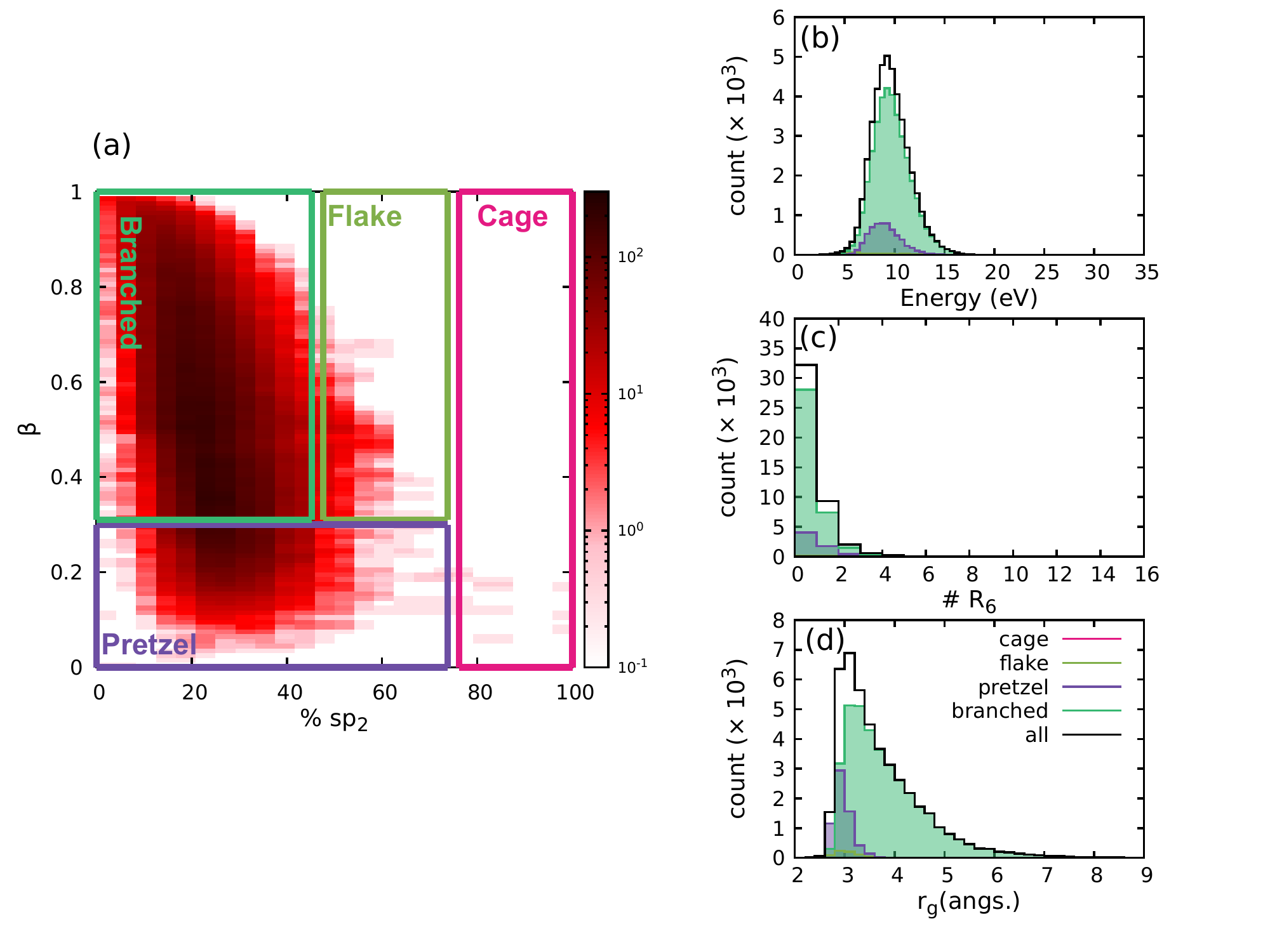} \\
                    \caption{Distributions of isomers of \cbn{24}  obtained from the PTMD/REBO  (i) and GA/DFTB structures (ii). Left: 2D distribution as a function of the sp$^{2}$ hybridization percentage (\%sp$^2$ and the asphericity parameter $\beta$ (a). Right: 1D distributions as a function of  energy (b), number of 6-carbon rings ($\# R_6$) (c) , and gyration radius $r_g$ (d).}
        \label{fig:dist_c24}
    \end{figure}

\begin{figure}
    \centering
    \begin{tabular}{cccc}
        \includegraphics[width=4cm]{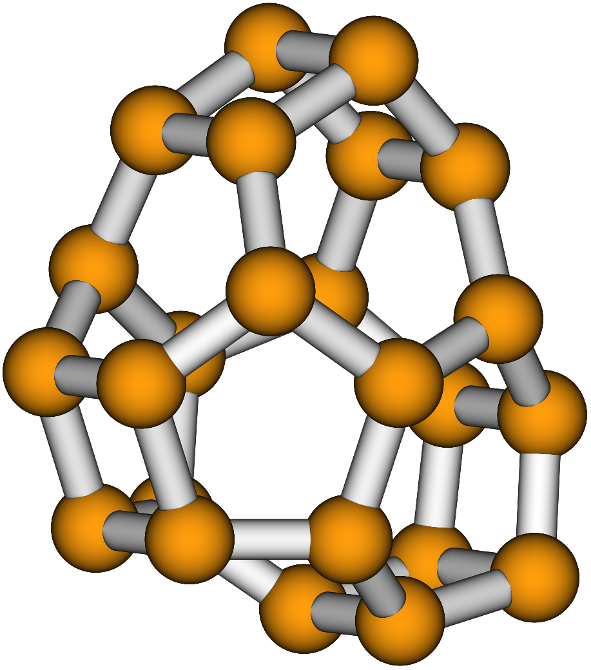} & \includegraphics[width=4cm]{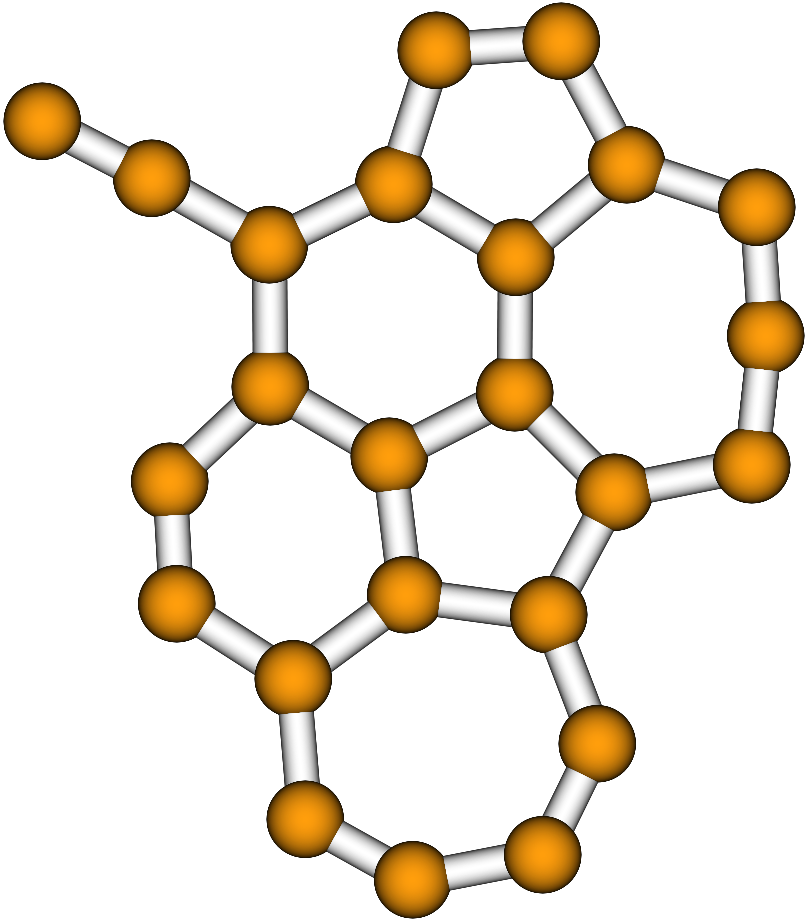} & \includegraphics[width=4cm]{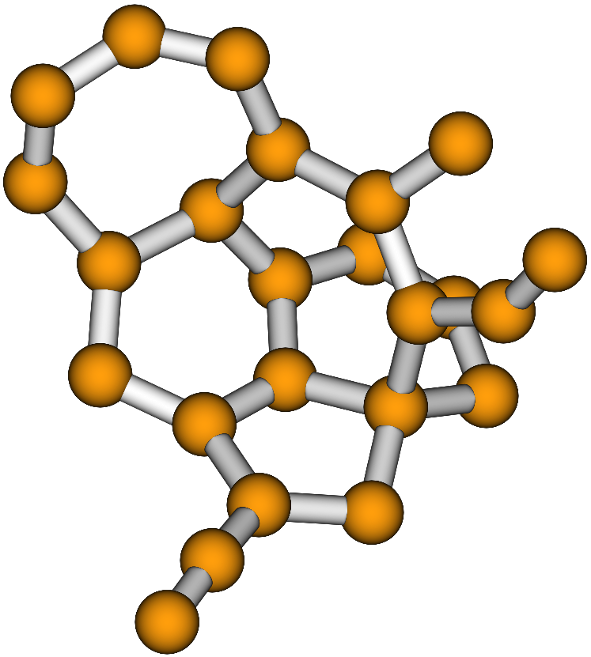} & \includegraphics[width=4cm]{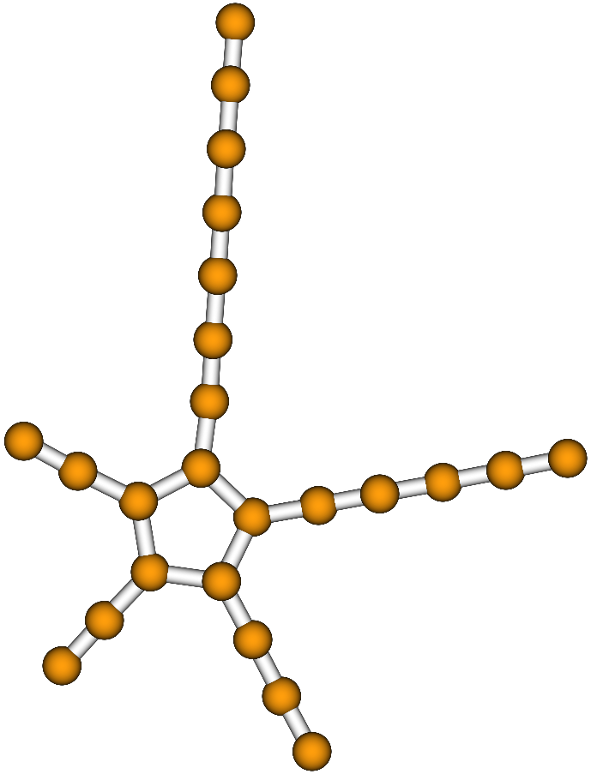}\\
        cage & flake & pretzel & branched
    \end{tabular}
    \caption{Examples of geometry per family for \cbn{24}. These correspond to the average of $\beta$ and $sp^2$ values for each family.}
    \label{fig:geom_c24}
\end{figure}

In the present work, we finally generated 15173 isomers for \cbn{24}. 
In Figure \ref{fig:dist_c24} (i, a) is reported the 2D distribution of isomers as a function of order parameters based on asphericity $\beta$ obtained from the Hill-Wheeler analysis, \cite{Hill1953} and percentage of sp$^{2}$ hybridized carbon that we determined in our previous work in order to define the isomers' families of \cbn{60} \cite{dubosq_mapping_2019-1}. This 2D distribution reported in Figure \ref{fig:dist_c24} (i, a) is continuous, therefore we cannot strictly define distinct families. However, we will use these parameters and the families' names to describe the evolution of the structures on the 2D map.  
The families' characteristics are reported in Table~\ref{tab:family_sp2} (a). The ones with the highest  sp$^{2}$ ratio are called cages.
Among the least spherical isomers ($\beta > 0.3$), those with a still high but lower sp$^{2}$ ratio are called flakes while more disordered isomers (fewer sp$^{2}$ carbon atoms) are named branched.
Isomers with more  spherical structures ($\beta < 0.3$) but sp$^{2}$ ratios below 75 \% are called pretzels. 
Following these definitions 7472 flakes, 7606 branched, 92 pretzels and 4 cages isomers were generated using the GA algorithm. Interestingly, dehydrogenated coronene was found as the most stable structure as expected\cite{kent2000c24,kosimov2010c24,Bonnin_2019} but we have not found the fullerene structure  which is not the most stable isomer but is in competition.\cite{manna2016c24} 
An example of isomer for each family is reported in Figure~\ref{fig:geom_c24}. \\

 Using the PTMD/REBO exploration followed by local SCC-DFTB optimization, 44341 isomers were found for \cbn{24}. The 2D distribution for these isomers is reported in Figure \ref{fig:dist_c24} (ii, a) and  11 cages, 714 flakes, 6307 pretzels, and 37309 branched structures were obtained (see Appendix in reference \cite{dubosq_mapping_2019-1}).  So branched structures are by far the most abundant, characterized by an extended gyration radius distribution (see Figure \ref{fig:dist_c24} (ii, d)). \\

 The energy profile shown in Figure~\ref{fig:dist_c24} (i,b)
 shows that the flakes generated by the GA algorithm are more stable than the branched isomers (maximum of the energy profiles shifted to the red), while in the case of PTMD/REBO simulations (Figure~\ref{fig:dist_c24} (ii,b)), practically all structures correspond to branched and pretzel structures that are less stable. 
 It thus seems that the GA algorithm  tends to explore more efficiently the lowest energy part of the PES for \cbn{24}.
 This better exploration of the lowest energy part is also reflected in the number of structures presenting 6-carbon rings, which should contribute to stabilize the system. When representing the number of structures as a function of the number of 6-carbon rings (Figure~\ref{fig:dist_c24} (c)), the maximum appears at around 2 rings for GA simulations while it is close to 0 for PTMD/REBO simulations. \\

\begin{table}[htbp!]
    \centering
    \begin{tabular}{c||c|c||c|c}
    & \multicolumn{2}{c||}{(a)} & \multicolumn{2}{c}{(b)} \\
        & \% sp$^{2}$ & $\beta$& $Rg (5,6)$& $\beta$  \\\hline
        Cages & > 100-75\% & 0.0-1.0& $\ge$ 9 & 0.0-1.0\\
        Flakes & 75-45\% & 0.3-1.0& 3-8 & 0.3-1.0\\
        Pretzels & 75-0\% & 0.0-0.3& 0-8 & 0.0-0.3\\
        Branched & 45-0\% & 0.3-1.0& 0-2 & 0.3-1.0\\\hline
    \end{tabular}
    \caption{Definitions of families based either on asphericity $\beta$ and percentage of sp$^{2}$ carbon atoms (a) or on  asphericity $\beta$ and number of 5 and 6-carbon rings $Rg (5,6)$ (b).}
    \label{tab:family_sp2}
\end{table}



\begin{figure}[htbp!]
        \centering
        \includegraphics[width=0.9\linewidth]{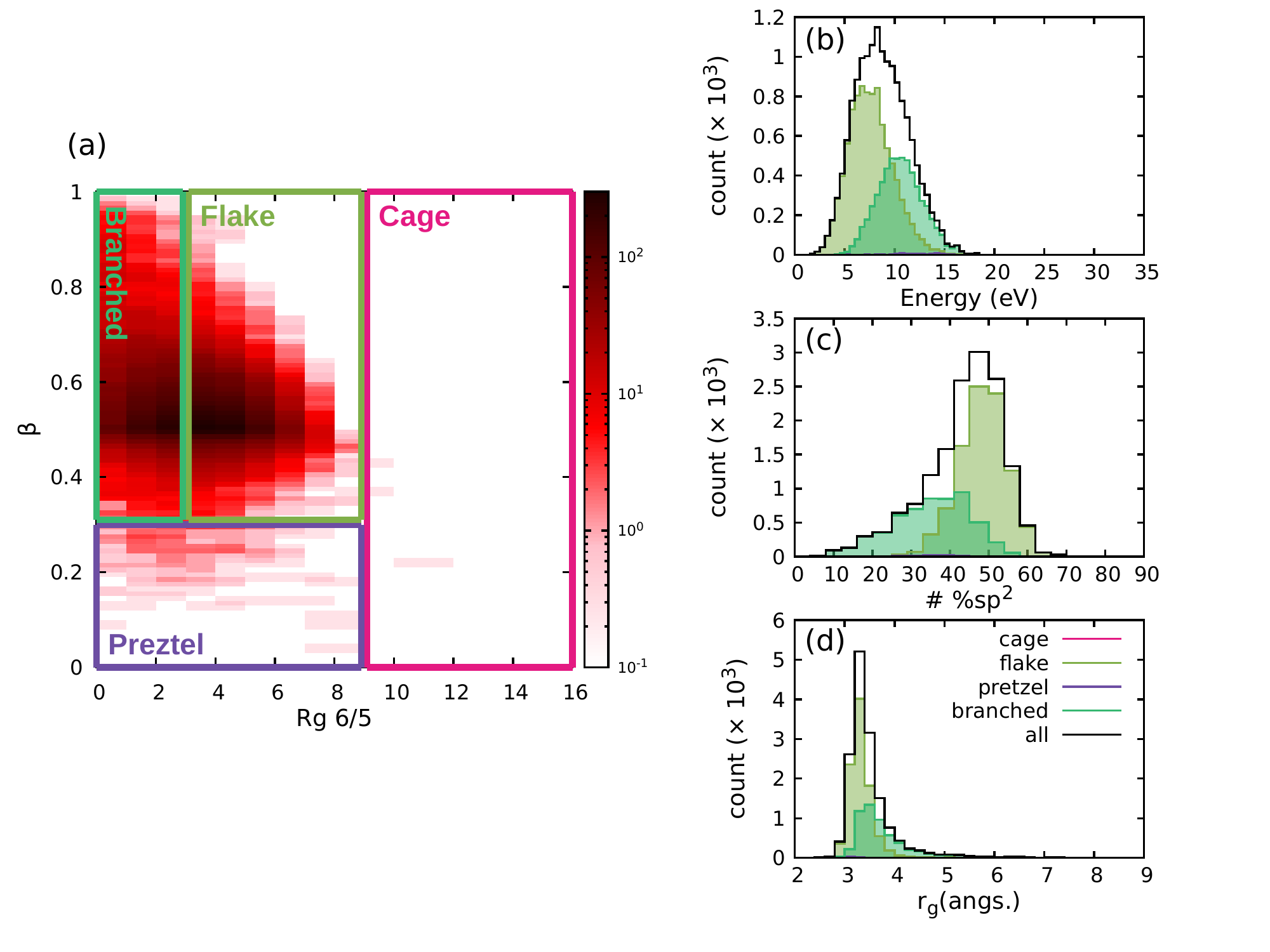}
        \caption{Distributions of \cbn{24} isomers. Left:  2D distribution as a function of $Rg (5,6)$ and $\beta$. Right: 1D distributions as a function of  energy (b), sp$^{2}$ hybridization ratio (c), and gyration radius $r_g$ (d).
}
        \label{fig:dist_c24_ring}
    \end{figure}

The sp$^{2}$ hybridization ratio appeared to be a suited parameter for pure carbon clusters. 
However, in the case of hydrogenated structures, the sp$^{2}$  parameter does not univocally describe a structural order. Indeed, in the case of pure carbon clusters, the carbon atoms which are not sp$^{2}$ are sp$^{1}$, whereas for higher hydrogenation ratios, sp$^{2}$ atoms can also be  replaced by sp$^{3}$ atoms. Therefore we decided to use the number of  5 or 6-carbon rings per structure (designed hereafter as $Rg (5,6)$) as a more general order parameter, combined with the asphericity parameter $\beta$. The distribution of isomers as a function of these two parameters ($Rg (5,6)$,  $\beta$) for \cbn{24} can be found in Figure~\ref{fig:dist_c24_ring} (a) and  the new definitions of the families are shown in Table~\ref{tab:family_sp2} (b).  \\

Using this definition, flakes are the most abundant (62\%, see Table~\ref{tab:nb_isom} 1st column), followed by branched (37\%). The highest density of flakes correspond to mostly planar structures ($\beta$ $\sim$ 0.5)  with more than two 5 or 6-carbon rings.  Branched structures possess less than two  5 or 6-carbon rings. Pretzels, more spherical, are far less abundant (0.6\%), and the most ordered structures (cages) 
 are rare (3 structures).  
In the following, families for hydrogenated clusters are refered to using the ($Rg (5,6)$,$\beta$) couple of order parameters.


\section{Results} \label{sec:res} 

\subsection{Fragmentation}

\begin{figure}[H]
        \centering
        \includegraphics[width=10.5cm]{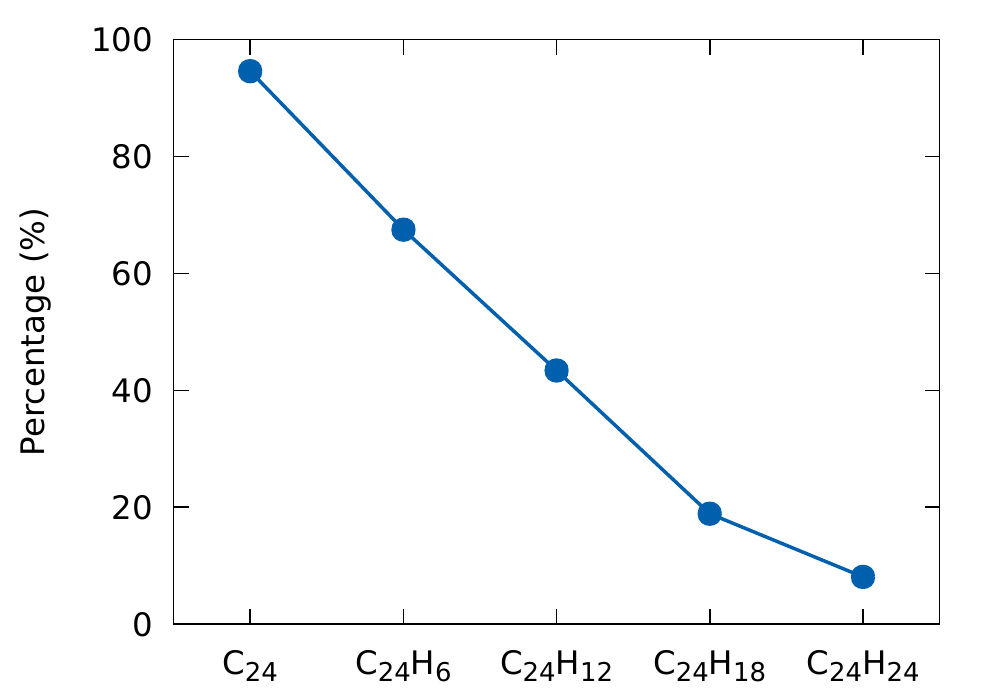}
        \caption{Percentage of non-fragmented structures of \cbnhm{24}{n} (\textit{n} = 0, 6, 12, 18, 24) obtained using the GA.}
        \label{fig:percnonfrag}
    \end{figure}

 As previously mentioned, during the optimization procedure some fragmented structures are removed from the final population. 
  As can be seen in Figure~\ref{fig:percnonfrag}, 
  the fraction of non-fragmented structures decreases when the hydrogenation rate increases. This highlights the difficulty in generating structures with a high  H/C ratio. 
 
\begin{figure}[H]
     \centering
     \begin{subfigure}[b]{0.49\textwidth}
         \centering
         \includegraphics[width=\textwidth]{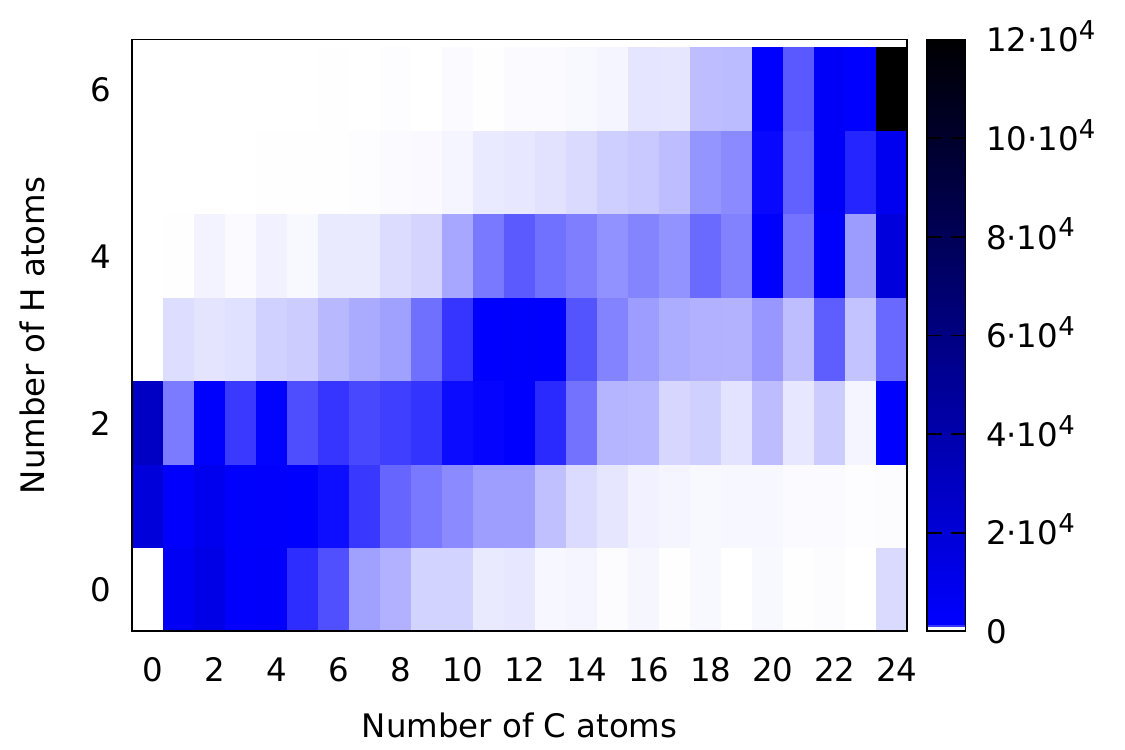}
         \caption{\cbnhm{24}{6}}
         \label{fig:allfragments_c24h6}
     \end{subfigure}
     \begin{subfigure}[b]{0.49\textwidth}
         \centering
         \includegraphics[width=\textwidth]{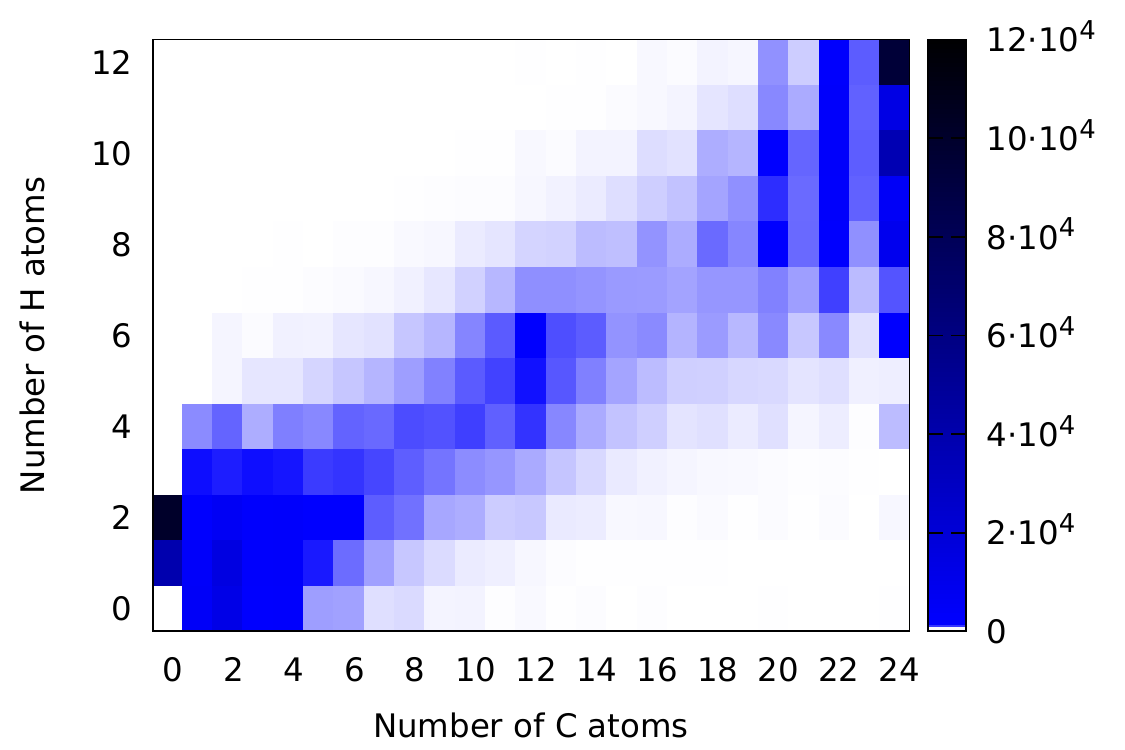}
         \caption{\cbnhm{24}{12}}
         \label{fig:allfragments_c24h12}
     \end{subfigure}
     \begin{subfigure}[b]{0.49\textwidth}
         \centering
         \includegraphics[width=\textwidth]{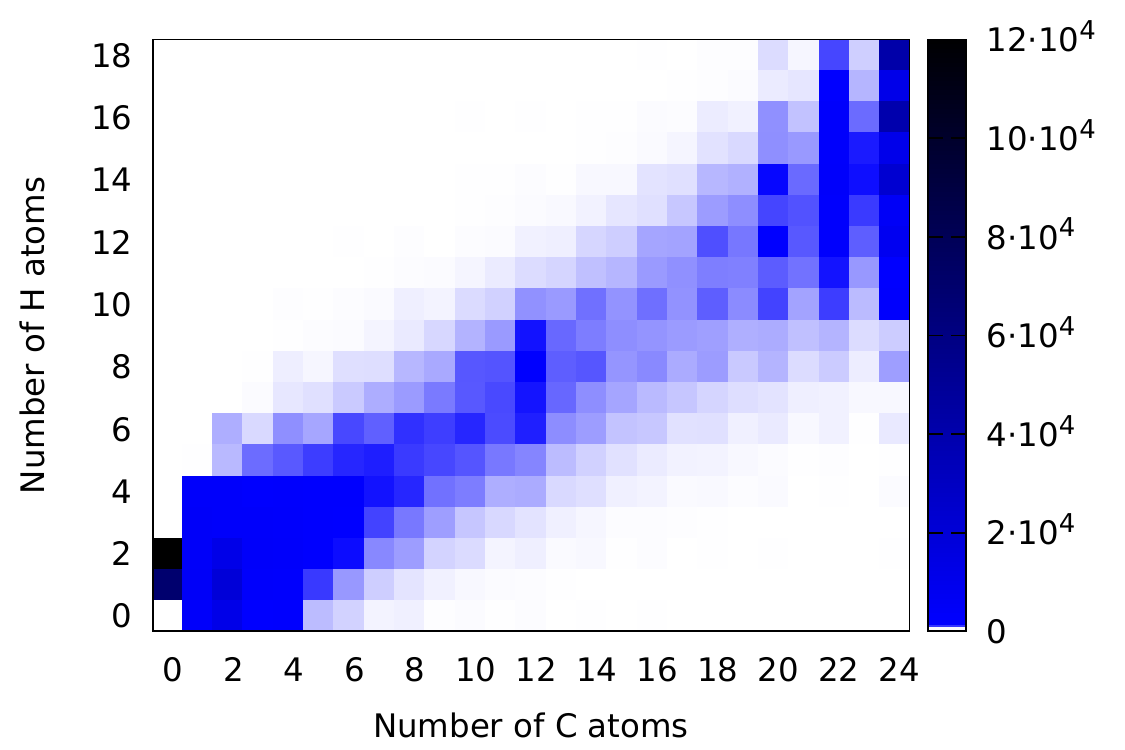}
         \caption{\cbnhm{24}{18}}
         \label{fig:allfragments_c24h18}
     \end{subfigure}
     \begin{subfigure}[b]{0.49\textwidth}
         \centering
         \includegraphics[width=\textwidth]{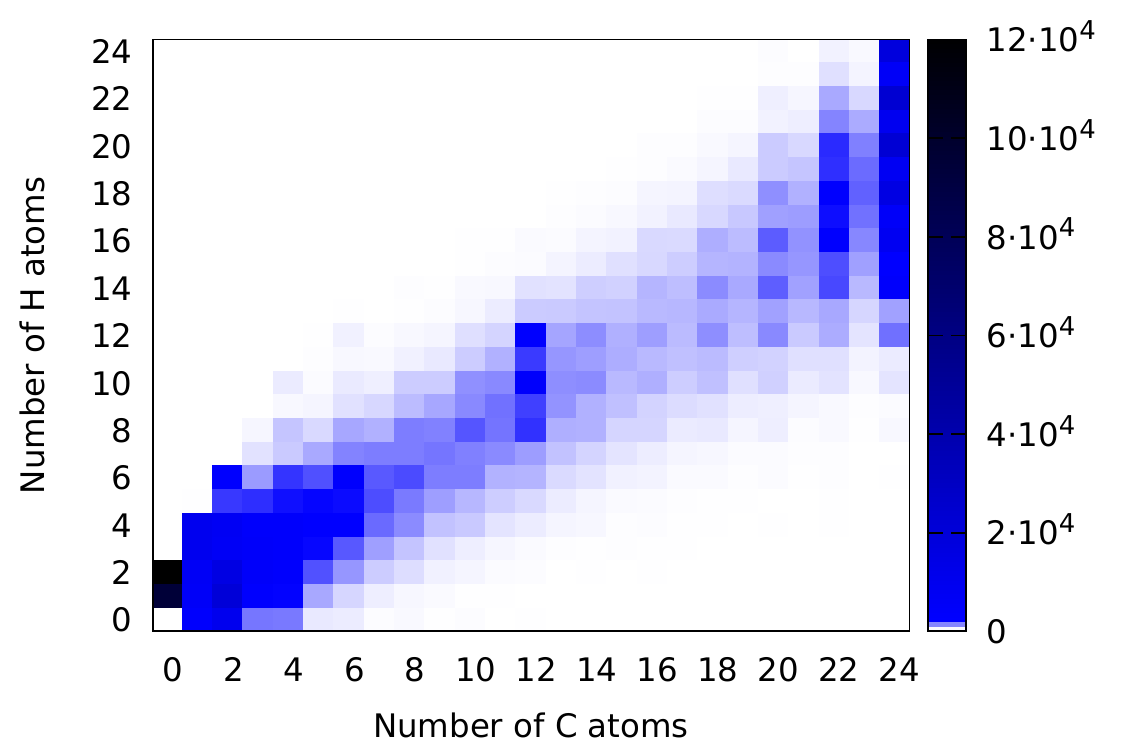}
         \caption{\cbnhm{24}{24}}
         \label{fig:allfragments_c24h24}
     \end{subfigure}
        \caption{Distribution of fragments as a function of the number of C and H atoms for  \cbnhm{24}{6}(a) , \cbnhm{24}{12} (b), \cbnhm{24}{18} (c) and \cbnhm{24}{24} (d). }
        \label{fig:allfragments}
\end{figure}

To get insights into the nature of the fragments we  analyzed all structures obtained during simulations. The procedure consisting in counting, for a given hydrogenation degree \cbnhm{24}{n=6,12,18,24}, the fragments with a defined number of C and H atoms. Figure \ref{fig:allfragments} shows the distribution of the fragments for each hydrogenation rate. \\

Interestingly, as indicated by the shapes of the distributions in Figure~\ref{fig:allfragments}, the H/C ratio of the fragmented structures remains similar to the initial set. 
We observe three areas with a higher number of fragments. The first one corresponds to small molecules which must be linear or small rings such as a pentagon or a hexagon, since the maximum number of carbon atoms is 6. These molecules present also a few H atoms. The second one corresponds to structures with a high number of C atoms mainly 24, 22 and 20.  The last area with a higher number of structures is centered at 12 C atoms and broadens with less H atoms (see for instance for \cbnhm{24}{6}). 
For simulations of  \cbnhm{24}{24}, the fragments \cbnhm{12}{12} and \cbnhm{12}{10} 
are especially stable, which again present a similar H/C ratio than the initial one. Interestingly, in particular for fragments with more than 12 C atoms, those with an even number of C atoms are major than those with an odd number of C atoms. For these fragments, an even number of H atoms is more favorable, leading to a stable closed-shell configuration. 
Many fragments observed in the simulations correspond to H atoms and \hbn{2} molecules and their number increases as the number of H atoms increases in the simulation. Although the color bar in the plots is set to a maximum of 120000  in all panels to simplify comparison,  the number of \hbn{2} molecules is higher than that one in some cases. We have obtained 28366 \hbn{2} molecules for \cbnhm{24}{6}, 100539 for \cbnhm{24}{12}, 232583  for \cbnhm{24}{18} and 410496  for \cbnhm{24}{24}. 
 The number of isolated H atoms is always smaller but it also increases with more H atoms in the simulation. The number of \hbn{2} molecules found is larger than the number of non-fragmented structures for \cbnhm{24}{12}, \cbnhm{24}{18} and \cbnhm{24}{24} but lower than the number of non-fragmented structures for \cbnhm{24}{6}. \\
Finally, the fragment analysis shows that the final number of stable structures  decreases when the hydrogenation degree increases and that fragments with even numbers of carbon and hydrogens appear to be favored.

\subsection{Evolution of  structural diversity as a function of hydrogen ratio}

\subsubsection{Overview}
\begin{figure}
    \centering
    \includegraphics[width=0.9\linewidth]{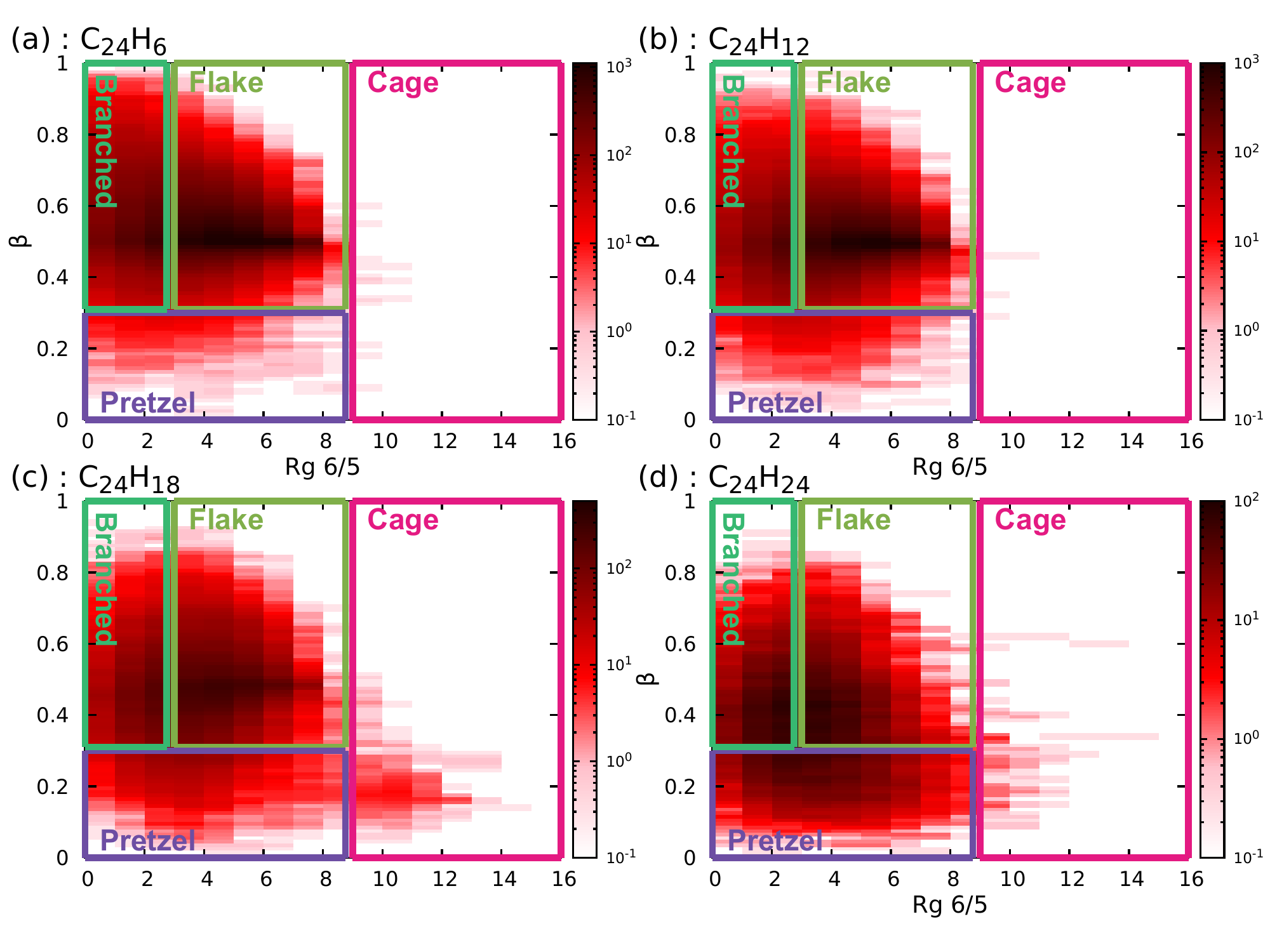}
    \caption{2D distributions of \cbnhm{24}{6} (a), \cbnhm{24}{12} (b), \cbnhm{24}{18} (c) and \cbnhm{24}{24} (d) isomers as a function of the order parameters $Rg (5,6)$ and $\beta$.}
    \label{fig:hydro_map}
\end{figure}

\begin{figure}
    \centering
    \begin{tabular}{cccc}
        \includegraphics[width=3cm]{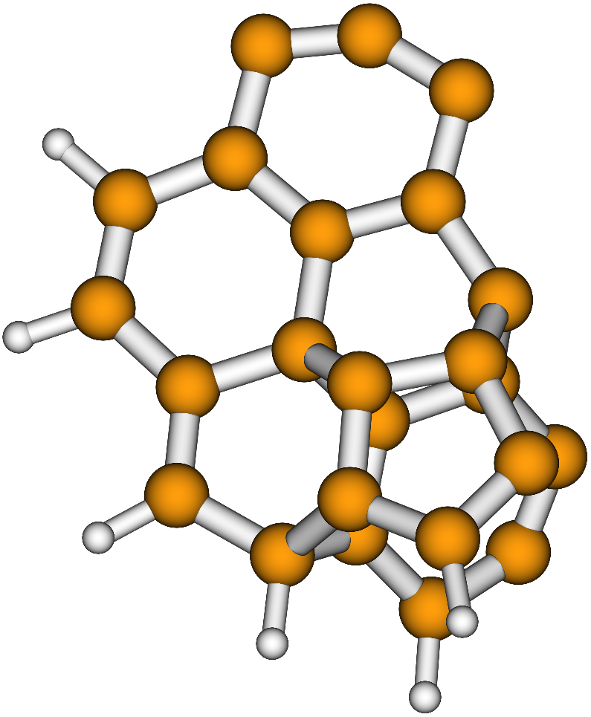} & \includegraphics[width=3cm]{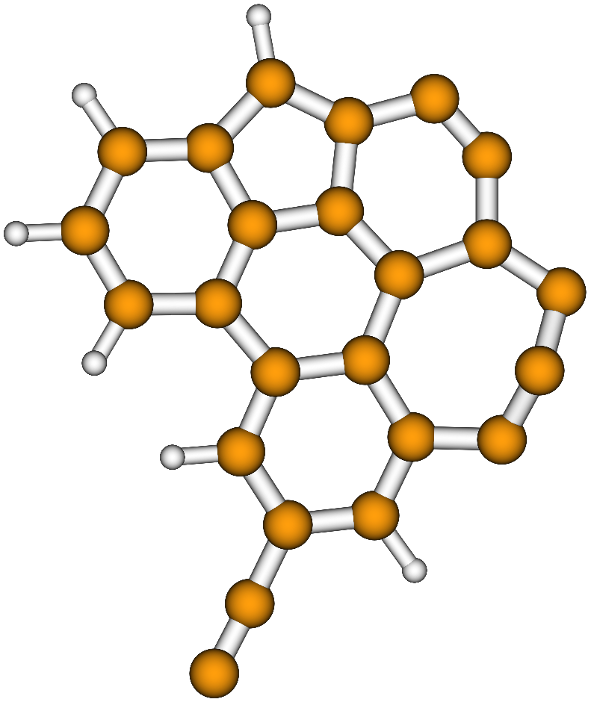} & \includegraphics[width=3cm]{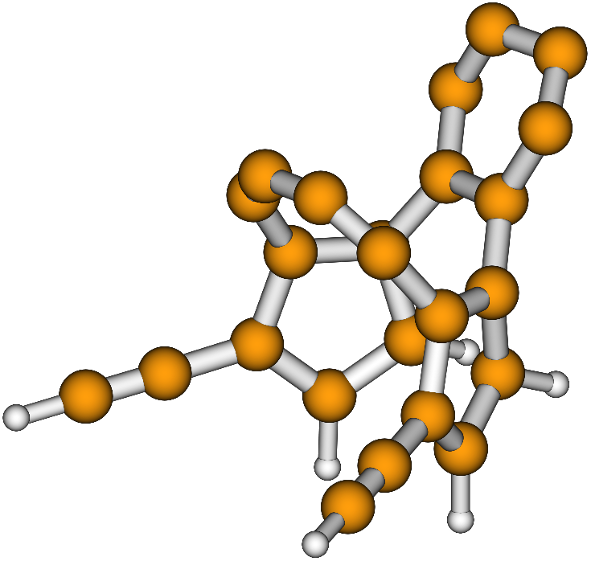} & \includegraphics[width=3cm]{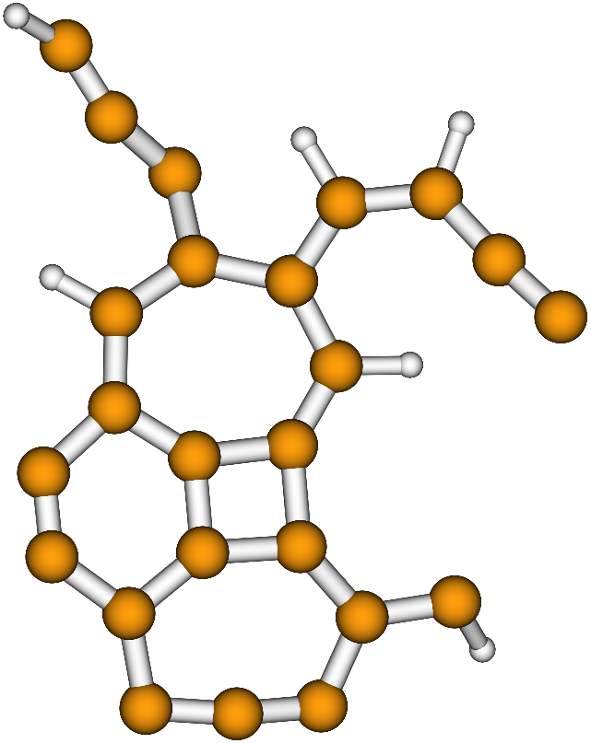}\\
        (a) & \cbnhm{24}{6} && \\
       &&& \\
        \includegraphics[width=3cm]{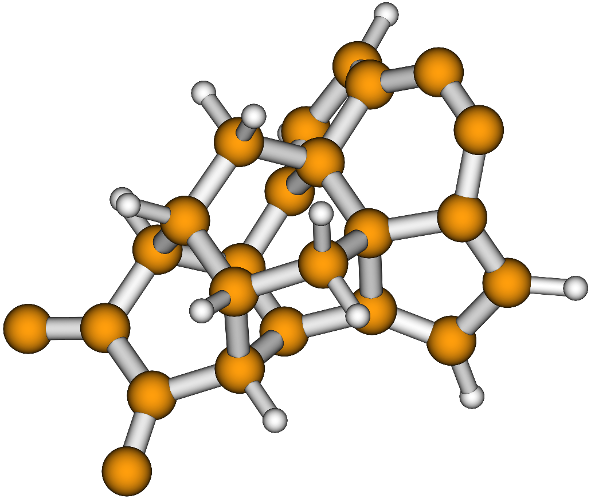} & \includegraphics[width=3cm]{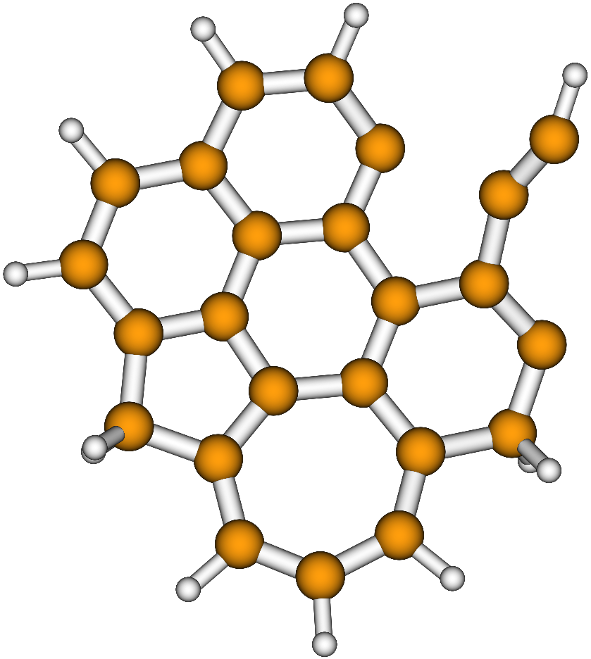} & \includegraphics[width=3cm]{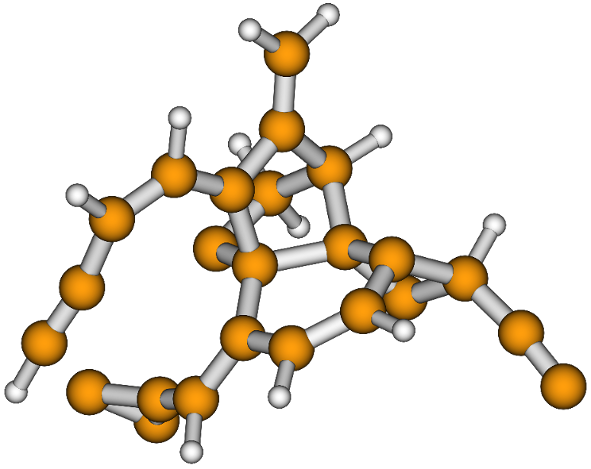} & \includegraphics[width=3cm]{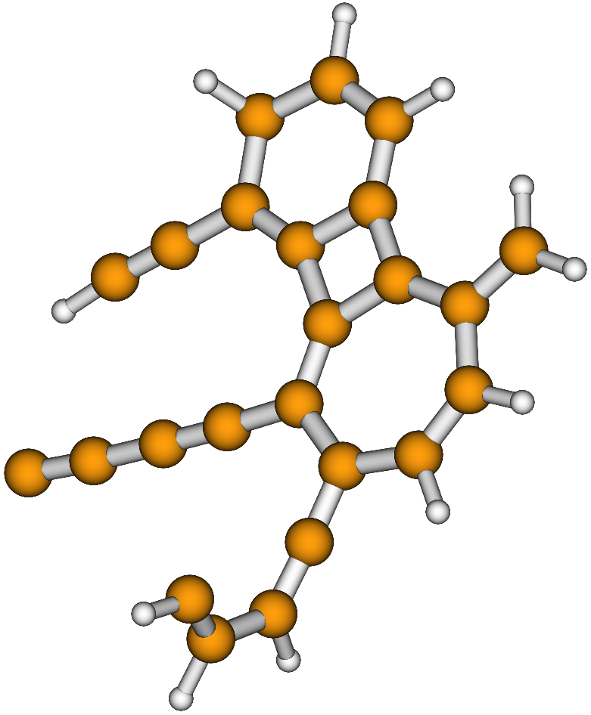}\\
         (b) & \cbnhm{24}{12}&& \\
      &&& \\
        \includegraphics[width=3cm]{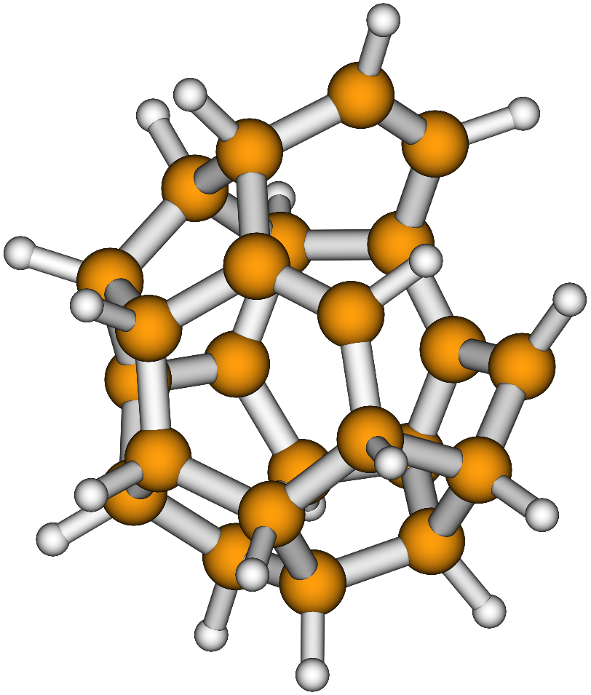} & \includegraphics[width=3cm]{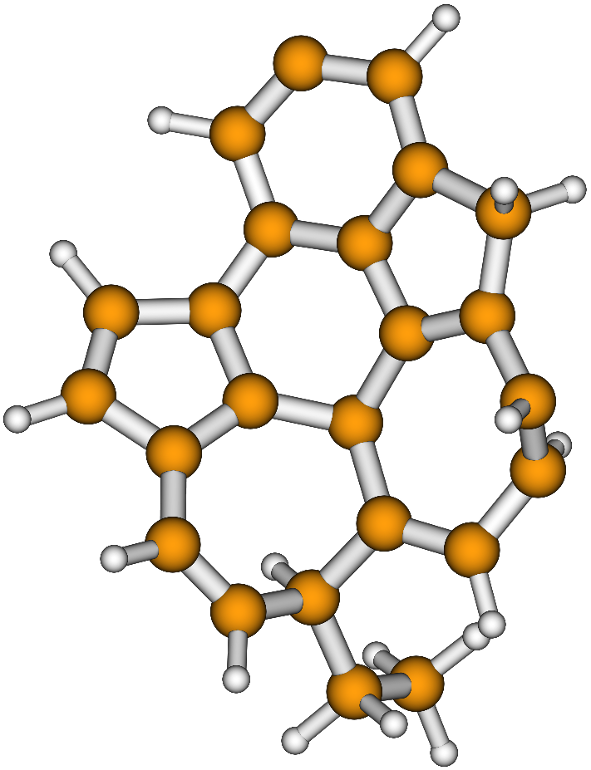} & \includegraphics[width=3cm]{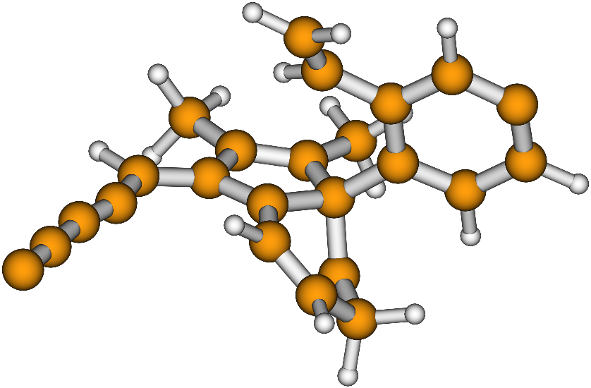} & \includegraphics[width=3cm]{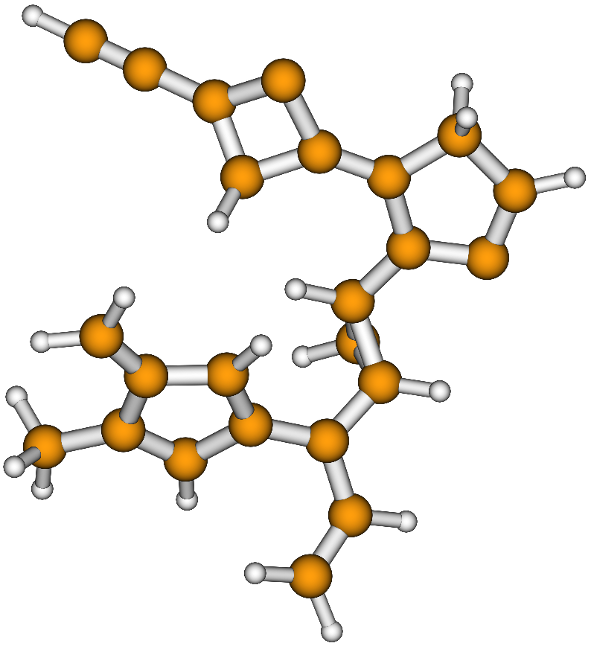}\\
(c) & \cbnhm{24}{18}&&  \\
&&& \\
 \includegraphics[width=3cm]{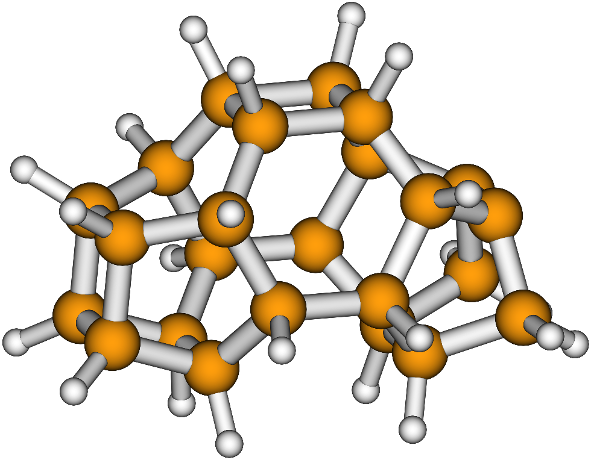} & \includegraphics[width=3cm]{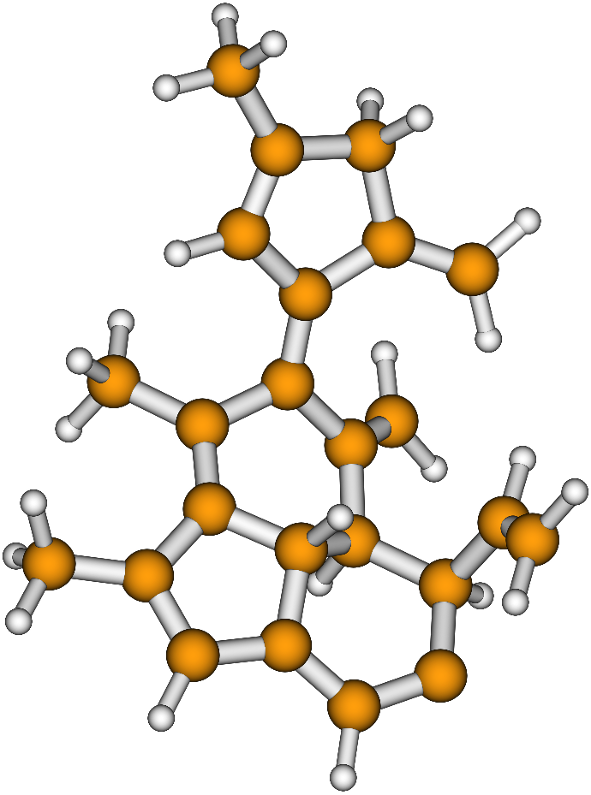} & \includegraphics[width=3cm]{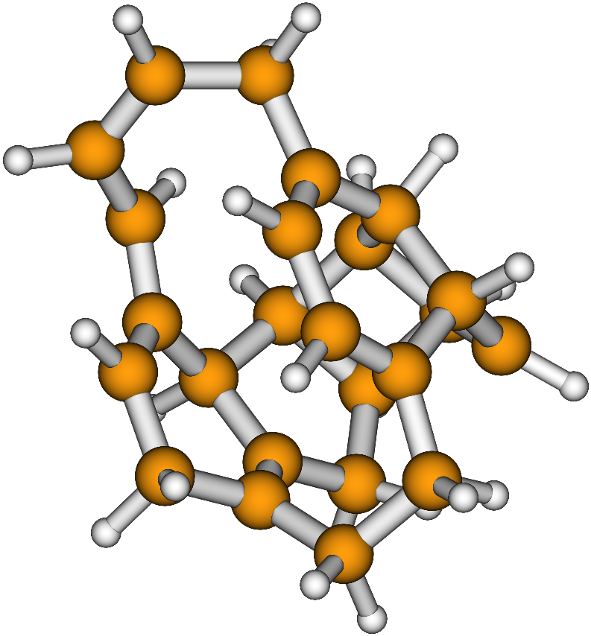} & \includegraphics[width=3cm]{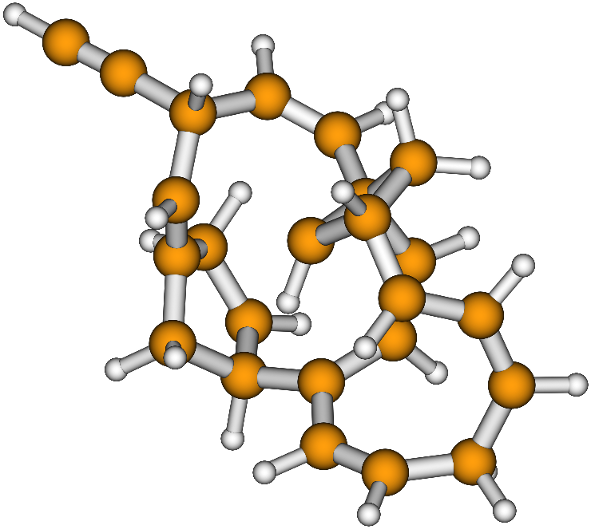}\\
(d) & \cbnhm{24}{24}&& \\
&&& \\
        cage & flake & pretzel & branched \\
    \end{tabular}
    \caption{Examples of isomers for each  family for \cbnhm{24}{6,12,18,24}.}
    \label{fig:geom_c24hn}
\end{figure}

\begin{figure}
    \centering
    \includegraphics[width=0.9\linewidth]{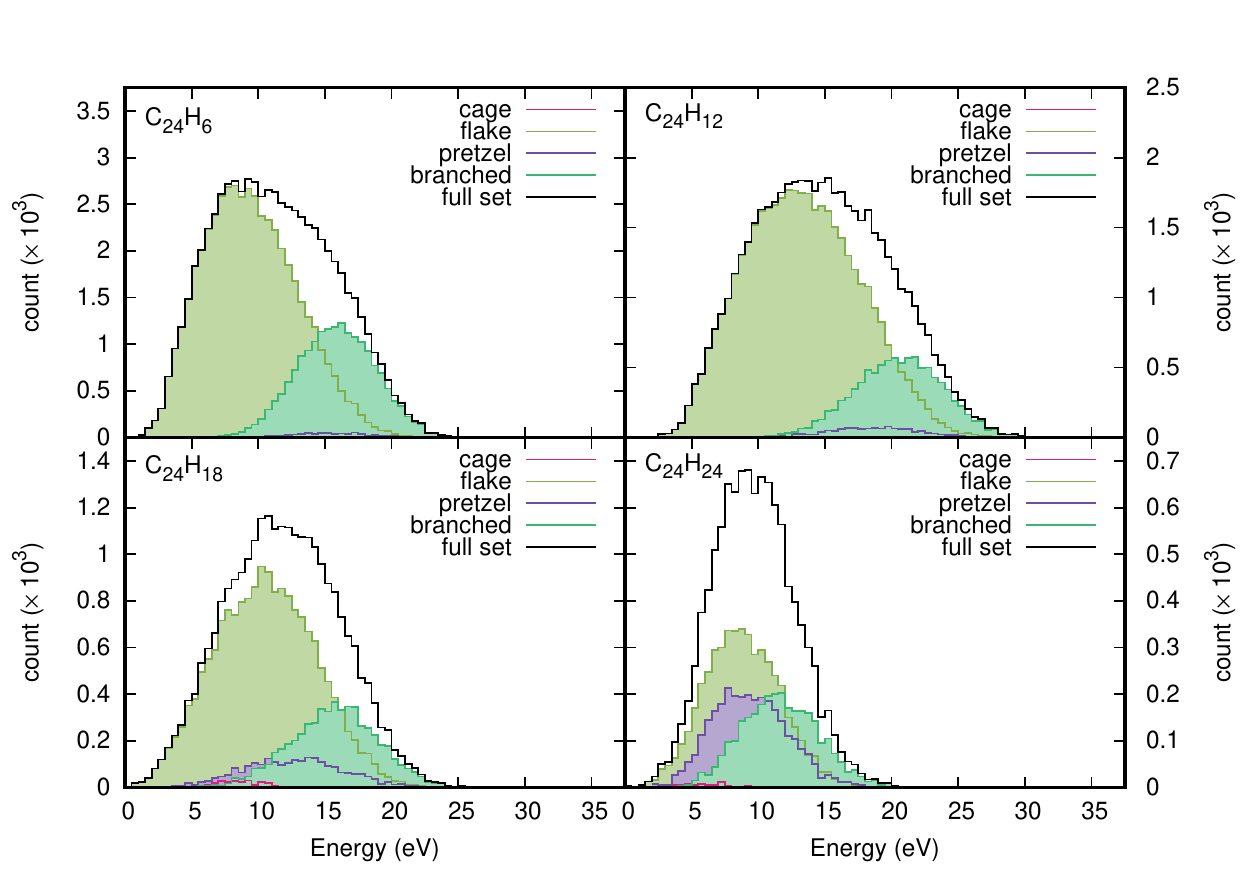}
    \caption{Distributions of \cbnhm{24}{6} (a), \cbnhm{24}{12} (b), \cbnhm{24}{18} (c) and \cbnhm{24}{24} (d) isomers as a function of energy}
    \label{fig:nrj}
\end{figure}

The total number of non fragmented and non redundant structures for \cbnhm{24}{n=6,12,18,24} 
as well as isomers per family, are reported in Table~\ref{tab:nb_isom}. 
We must specify here that the total number of isomers found by this technique remains orders of magnitude smaller that the number of reasonable structures that can be formed for each stoichiometry, all the more as the number of hydrogen atoms increases. This is probably due to the sampling limitation inherent to the algorithm, and certainly to the increase of the fragmentation efficiency when the hydrogenation rate increases (Figure 4). The 2D distributions of these isomers as a function of the order parameters  $\beta$ and  $Rg (5,6)$ for all hydrogenation degrees are reported in Figure~\ref{fig:hydro_map}, with examples of isomers  for each family in Figure~\ref{fig:geom_c24hn}.  The distributions of isomers  as a function of energy are reported in Figure~\ref{fig:nrj}.  \\

\begin{table}
    \centering
    \begin{tabular}{c|ccccc|}
              n & 0& 6 &  12 & 18 & 24 \\
              \hline
             total  & 15173 & 68896 & 52213 & 26019 & 10894 \\
              flakes & 9449 (62\%) & 49989 (73\%) & 41264 (79\%) & 17889 (69\%) & 4924 (45\%) \\
              branched & 5626 (37\%) & 18191 (26\%) & 9722 (19\%) & 5578 (21\%) & 2889 (27\%) \\
              pretzel & 95(0.6\%) & 704 (1 \%) & 1224 (2.3\%) & 2293 (9\%) & 3009 (28\%) \\
              cage & 3 & 12 & 5 & 259 (1\%) & 72 \\
              \hline
    \end{tabular}
    \caption{Total number of non fragmented and non redundant structures and numbers of isomers per family obtained with the GA algorithm for all C$_{24}$H$_n$ systems. 
}
    \label{tab:nb_isom}
\end{table}


As can be seen in Table~\ref{tab:nb_isom}, up to n$_H$=18, 
 flakes 
 are major (62\% to 79\%), followed by branched (19\% to 37\%). In terms of stability, flakes are also the most stable structures (see Figure~\ref{fig:nrj}), followed by pretzels and branched.  When n$_H$=24, 
 the flakes' population remains major but to lesser extent (45\%) while the pretzel's population increases to become similar to the branched population (28\% {\it vs} 27\%). 
 Interestingly, flakes and pretzels's populations have similar energy distributions (Figure~\ref{fig:nrj}) while branched structures remain higher in energy. For all hydrogenation degrees, cages, i.e. the most ordered and spherical isomers, represent a tiny fraction of the population. \\ 
 


\begin{figure}[htbp!]
    \centering
    \begin{tabular}{cccc}
        \includegraphics[width=4cm]{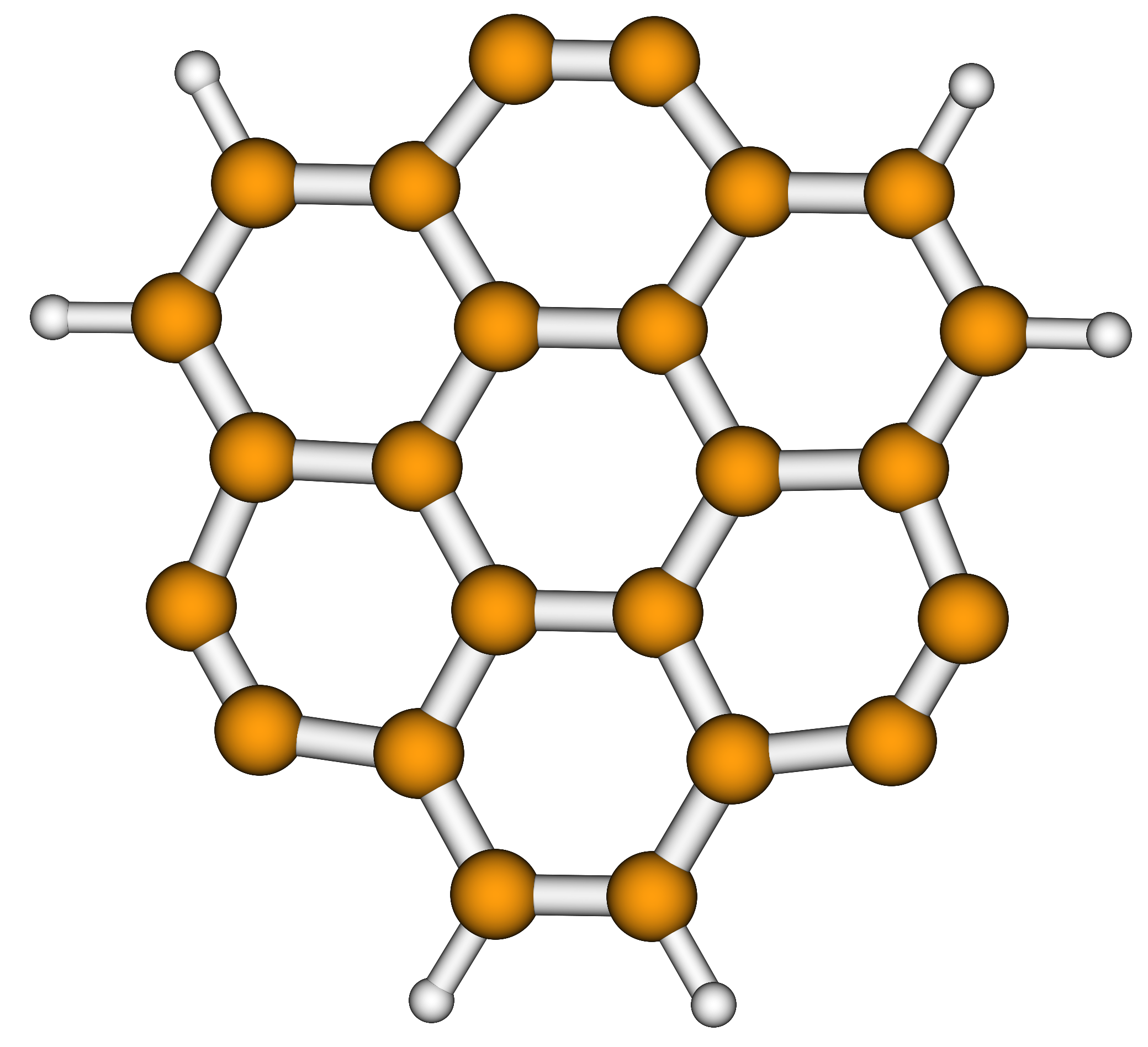} & \includegraphics[width=4cm]{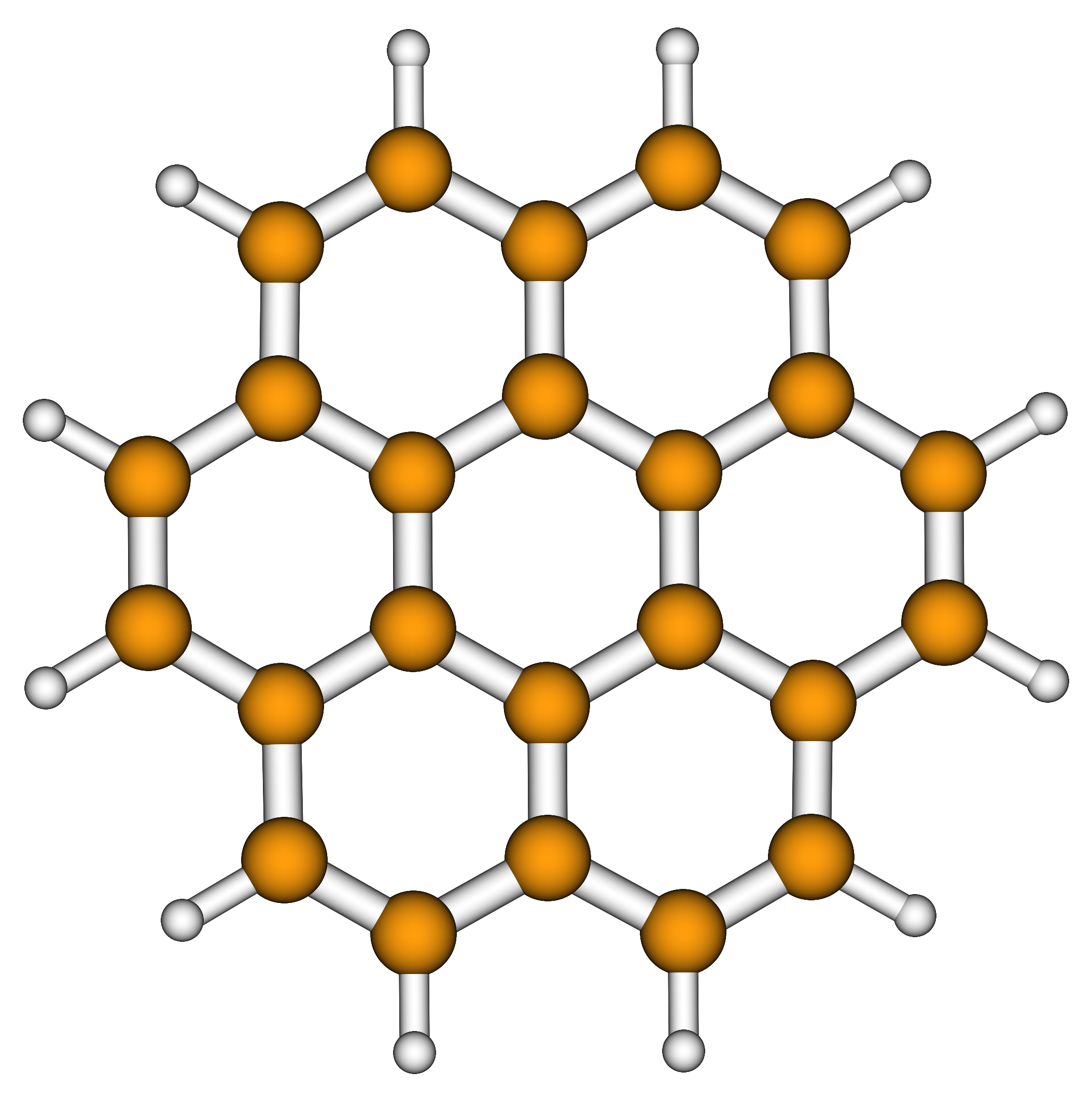} & \includegraphics[width=4.4cm]{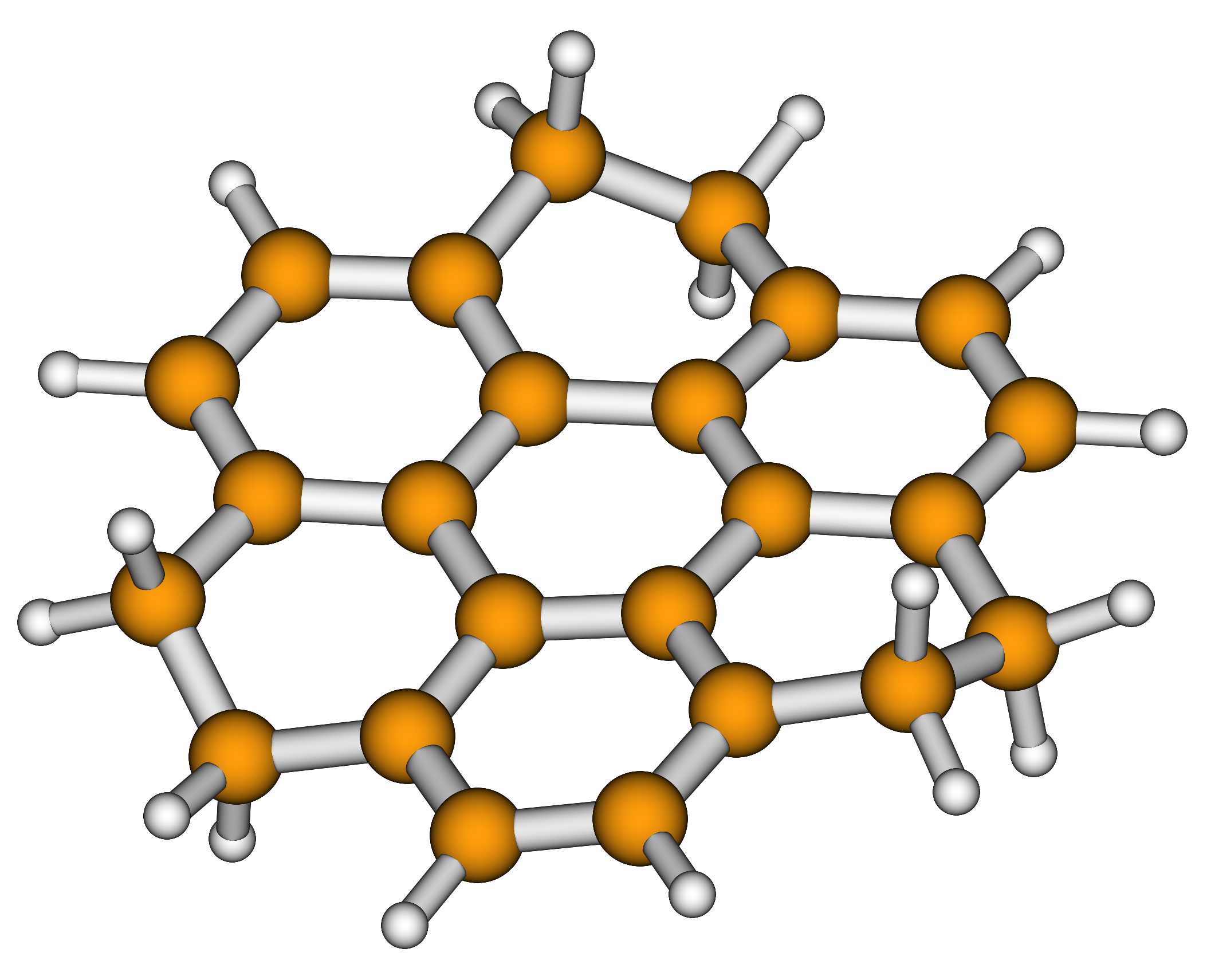} & \includegraphics[width=3.6cm]{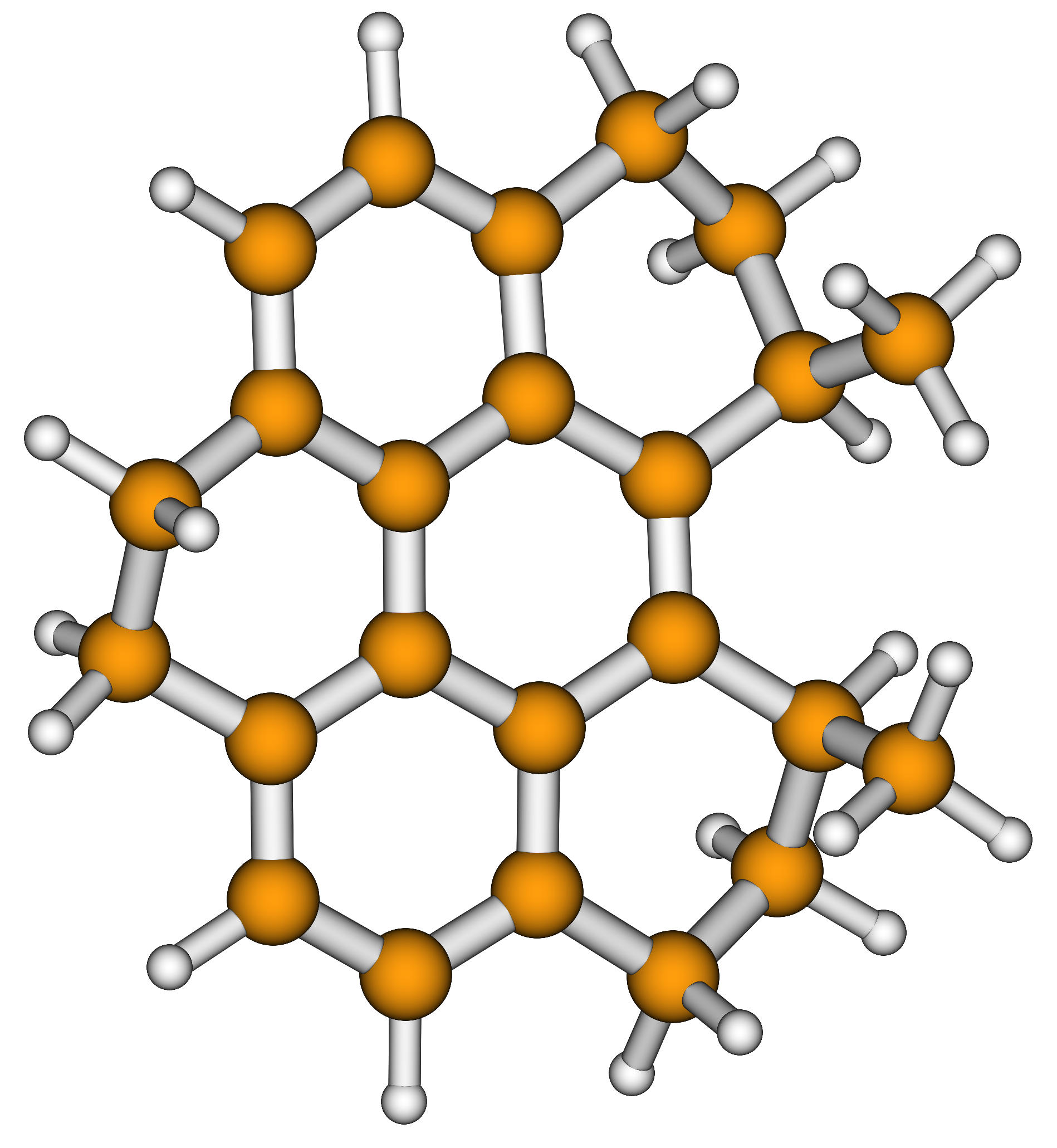}\\
        \ \cbnhm{24}{6} & \cbnhm{24}{12} & \cbnhm{24}{18} & \cbnhm{24}{24}
    \end{tabular}
    \caption{Most stable isomers obtained with GA for \cbnhm{24}{n} (n = 6, 12, 18 and 24). }
    \label{fig:geom_moststable}
\end{figure}

It is interesting to underline that the most stable structures found for hydrogenated clusters correspond to derivatives of coronene (flakes with seven hexagonal rings) except for \cbnhm{24}{24} where one of the rings has broken in favor of two methyl groups (see Figure \ref{fig:geom_moststable}). As mentioned hereabove, in previous works \cite{kent2000c24,kosimov2010c24,Bonnin_2019}, the most stable isomer found for \cbn{24} was dehydrogenated coronene except when using explicitly correlated methods where the cage structure is found to be a few kcal/mol more stable than dehydrocoronene.\cite{manna2016c24} For \cbnhm{24}{6}, three isomers are found very close in energy (less than 0.01 eV) which correspond to isomers in which the hydrogen atoms are placed by pairs. These dehydrogenated coronene derivatives containing hydrogen atoms placed by pairs have been previously studied as they are more likely to be found in interstellar conditions.\cite{malloci2008dehydrogenated} Coronene is the most stable isomer of \cbnhm{24}{12} as reported by Kosimov \textit{et al.}\cite{kosimov2010c24} being 1.33 eV more stable than the next isomer found by GA. Finally, it is worth noticing that the most stable isomer of \cbnhm{24}{18} corresponds to the most stable isomer of hydrogenated coronene with 6 additional H atoms\cite{Pla2020PCCP} 
whereas the calculation of the most stable isomer of hydrogenated coronene with 12 additional H atoms\cite{Pla2020PCCP} calculated at SCC-DFTB level reveals that it is 0.31 eV higher in energy than the most stable one obtained in the present work.

\subsubsection{Structural evolution as a function of $n_H/n_C$ }
In this section, we provide a detailed description of the families' characteristics  by ascending order of hydrogen rate in \cbnhm{24}{n}. 
The results are discussed in the light of Table~\ref{tab:nb_isom}, Figures~\ref{fig:hydro_map} and ~\ref{fig:nrj}, as well as additional Figures in the Supporting Information,  that represent the distribution of isomers for each family as a function of the sp$^2$ ratio (Figure~S1)

 and the gyration radius $r_g$ (Figure~S2).  


\paragraph{$\bullet$\cbnhm{24}{6}}~\\

    
 With respect to \cbn{24}, a larger fraction of flakes structures was obtained for $n_H/n_C=1/4$ (73\% vs 62\%) to the detriment of branched structures (26\% vs 37\%) showing that the presence of a few H atoms favors more ordered structures. 
The  energetics profile (Figure~\ref{fig:nrj}) is wider than for C$_{24}$ (Figure~\ref{fig:dist_c24_ring},)
indicating a spreading of the structures with less degeneracy. Besides, it is clearly asymmetric with a sharp red increase and a loose blue tail. The density of structures is located at higher energy (9.07 \si{\electronvolt} vs  8.12 \si{\electronvolt} in the case of \cbn{24}). 
Regarding the electronic and structural properties, adding 6 hydrogen atoms to C$_{24}$ leads to an increase of  the sp$^{2}$ carbon ratio maximum (68\% for \cbnhm{24}{6} vs 48\% for \cbn{24}).
 Concerning the gyration radius, it  shifts the positions of the maxima toward smaller values (3.29\,\AA ~for \cbnhm{24}{6} vs 3.31\,\AA ~for \cbn{24}).

\paragraph{$\bullet$\cbnhm{24}{12}}~\\

  
When $n_H/n_C=1/2$, 
a slightly larger fraction of flakes structures is obtained with respect to $n_H/n_C=1/4$ (79\% vs 73\%), the maximum density of structures remaining concentrated in the same region (for a $\beta$ $\sim$ 0.5 and $Rg(5,6)$ between 3 and 9). The branched population slightly decreases (26\% to 19\%) whereas the pretzel population slightly increases (1\% to 2.3\%). Regarding the energy profile, it continues widening, with structures reaching 30\,eV above the most stable isomer. 
The sp$^{2}$ fraction maximum increases up to reaching its maximum among all  $n_H/n_C$ ratios (85\%)  while the gyration radius slightly decrease (3.26\,\AA).



\paragraph{$\bullet$\cbnhm{24}{18}}~\\   
    

For $n_H/n_C=3/4$, 
the fraction of flakes  structures decreases (69\% vs 79\% for $n_H/n_C=1/2$), that of  branched structures remains steady (21\% vs 19\% for $n_H/n_C=1/2$), while the fraction of pretzels structures increases (9\% vs 2.3\% for $n_H/n_C=1/2$), as well as that of cages (1\%) which now appears as a spot on the 2D map (Figure~\ref{fig:hydro_map} c)). Interestingly, pretzels and cages possess a more spherical character than flakes and branched isomers. Regarding the energy profile, it now tightens, all structures remaining less than 25 eV above the most stable isomer. The maximum of the \%sp$^2$ profile is now located at 77\%  while the gyration radius maximum continues to decrease down to 3.22\,\AA. It can be understood as increasing the number of H atoms leads to the existence of a larger number of more spherical and hence more compact isomers.





\paragraph{$\bullet$\cbnhm{24}{24}}~\\
   

 When $n_H/n_C=1$, 
 the structural diversity presents a drastic change :  the flakes population is still dominant (45\% vs 69\% for $n_H/n_C=3/4$) but the fraction of pretzel population increases (28\% vs 9\% for $n_H/n_C=3/4$). 
 The energy profiles of both populations become similar with the low energy tail corresponding to the most stable isomers. In line with the partial conclusions drawn for \cbnhm{24}{18}, increasing the $n_H/n_C$ ratio when it is higher than 1 leads to an increased populations of stable spherical isomers. 
The fraction of the branched population remains steady while that of the cage  decreases with respect to  $n_H/n_C=3/4$. 
The maximum of the \%sp$^2$ profile is now  down to  60\%, which is partly due to the presence of an increased fraction of  \%sp$^3$ atoms as reported in next subsection.  The positions of the gyration radius maximum is stabilized at around 3.21 $\AA$ corresponding again to a population increase of more spherical and more compact structures due to the high hydrogenation degree. 



\subsubsection{Evolution of hybridization and rings' distribution as a function of energy }
   
    In the following section, we propose an analysis of the evolution of carbon hybridization and rings' distributions as a function of energy for all families of all hydrogenation rates.
    
\begin{figure}[H]
     \centering
     \begin{subfigure}[b]{0.49\textwidth}
         \centering
         \includegraphics[width=\textwidth]{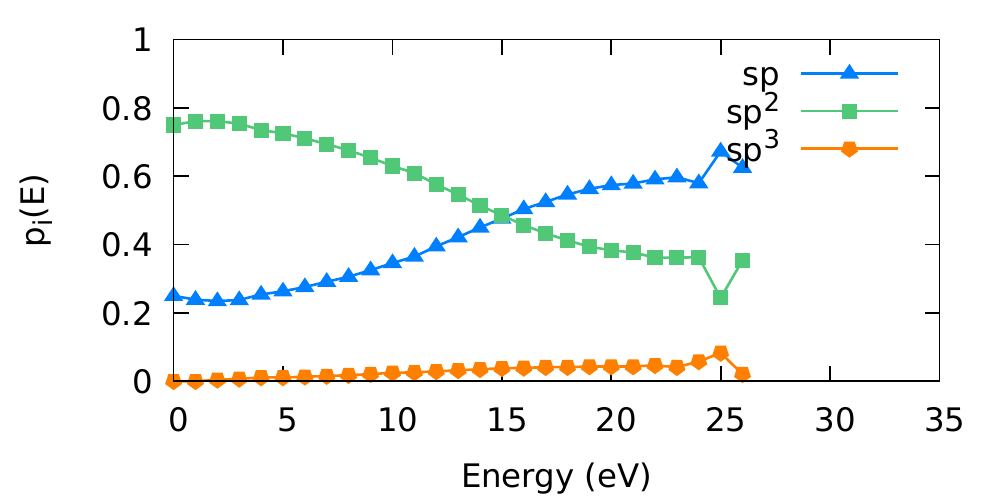}
         \caption{\cbnhm{24}{6}}
         \label{fig:hybrid_ener_c24h6}
     \end{subfigure}
     \begin{subfigure}[b]{0.49\textwidth}
         \centering
         \includegraphics[width=\textwidth]{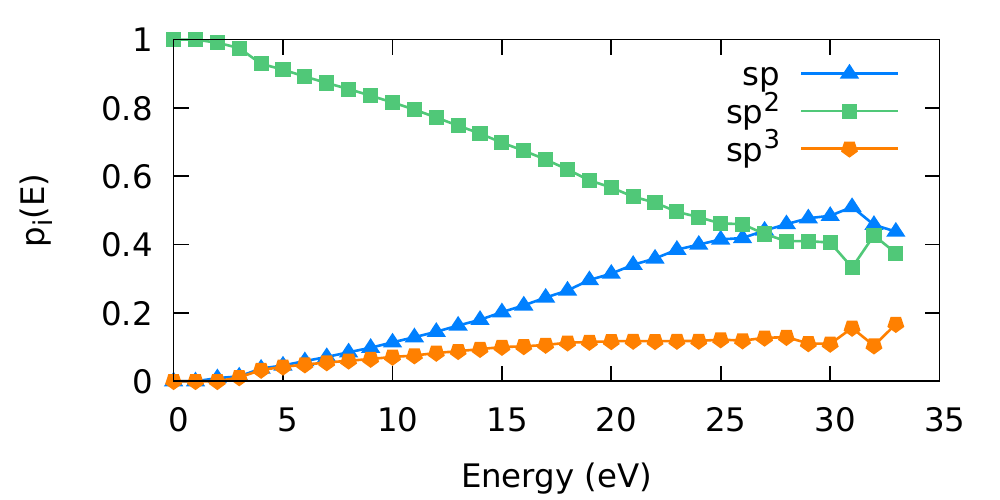}
         \caption{\cbnhm{24}{12}}
         \label{fig:hybrid_ener_c24h12}
     \end{subfigure}
     \begin{subfigure}[b]{0.49\textwidth}
         \centering
         \includegraphics[width=\textwidth]{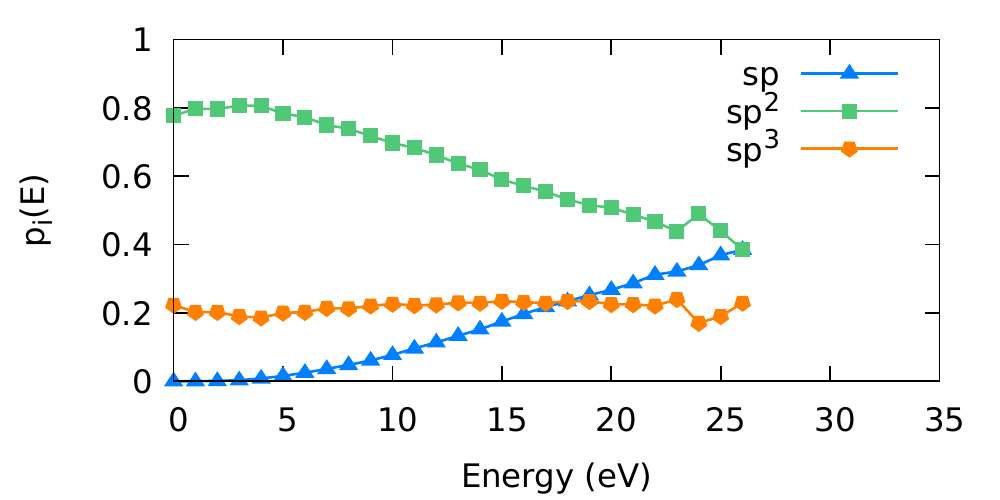}
         \caption{\cbnhm{24}{18}}
         \label{fig:hybrid_ener_c24h18}
     \end{subfigure}
     \begin{subfigure}[b]{0.49\textwidth}
         \centering
         \includegraphics[width=\textwidth]{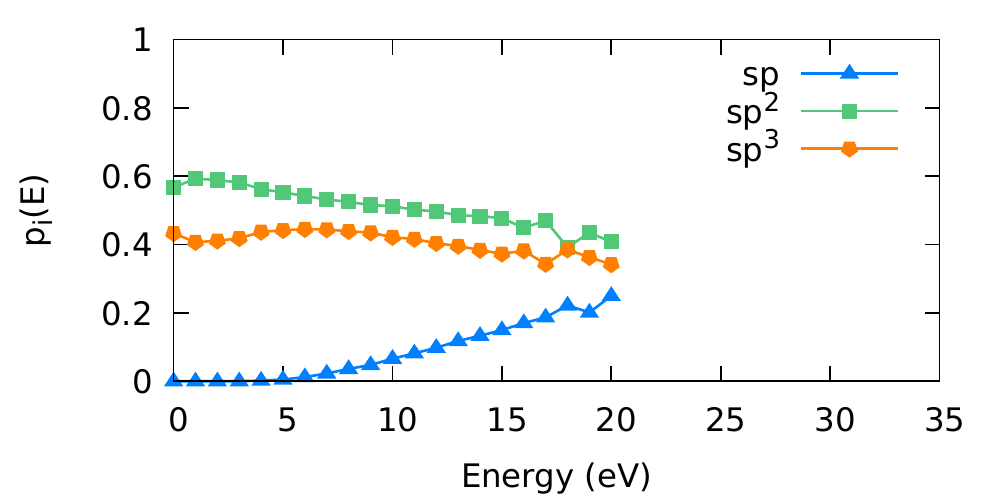}
         \caption{\cbnhm{24}{24}}
         \label{fig:hybrid_ener_c24h24}
     \end{subfigure}
        \caption{Evolution of the fraction of sp$^{i}$ (i=1-3) carbon atoms as a function of energy p$_i$(E)  for \cbnhm{24}{6} (a), \cbnhm{24}{12} (b), \cbnhm{24}{18} (c) and \cbnhm{24}{24} (d). The data points are obtained averaging the sp$^{i}$ carbon atoms' fraction for 1\,eV energy windows. This quantity becomes less valuable in the higher energy region (about 5 eV below the maximum energy) due to the low number of structures in these energy windows.}
        \label{fig:hybrid_vs_ener}
\end{figure}

\paragraph{$\bullet$Hybridization and stability}~\\
We first analyse the hybridization evolution based on Figure~\ref{fig:hybrid_vs_ener}. The sp$^{i}$  (i=1-3) character of the  C atoms was obtained using bond order as in our previous work \cite{dubosq_mapping_2019-1}. We however note that  the same results were obtained by simply counting the number of neighbouring atoms : a C atom is of sp$^{i}$ (i=1-3) character if it is covalently bonded  to i+1 atoms.  

As can be seen in Figure~\ref{fig:hybrid_vs_ener}, the  sp$^{2}$ fraction (p$_2$) is overall dominant  for all hydrogenation rates, which is consistent with the major presence of flakes structures. This  predominance is major for $n_H$=12 where the most stable  structures possess  100\% of sp$^{2}$ carbon atoms. This is not the case for the others: the sp$^{1}$ fraction (p$_1$) amounts to $\sim$0.25 for $n_H$=6, while the sp$^{3}$ fraction (p$_3$) is $\sim$ 0.22 and 0.43 for $n_H$=18 and 24, respectively. When energy increases, p$_2$ decreases, which is true for all hydrogenation rates but to various extents. The variation is maximum for $n_H$=12 (1 down to $\sim$ 0.38), minimum for $n_H$=24 (0.57 down to $\sim$0.41). sp$^2$ atoms are not major anymore for energies above 15 eV in the case of $n_H$=6 where p$_1$ becomes higher ($\sim$ 0.63  vs $\sim$ 0.35)). In the case of $n_H$=12,  p$_1$ and p$_2$ become equal for the very higher energy (above 27 eV). For $n_H$=18, p$_{3}$ is found to be steady along the whole energy range (0.18-0.24) while  p$_{1}$ increases from 0 to $\sim$ 0.39. In the case of $n_H$=24, the smallest variations are observed. p$_3$ also remains steady (about 0.40) while p$_{1}$ increases from 0.0 to 0.25 at the highest energy (20 eV).

\paragraph{$\bullet$Rings' distribution and stability}~\\

\begin{figure}[H]
     \centering
     \begin{subfigure}[b]{0.49\textwidth}
         \centering
         \includegraphics[width=\textwidth]{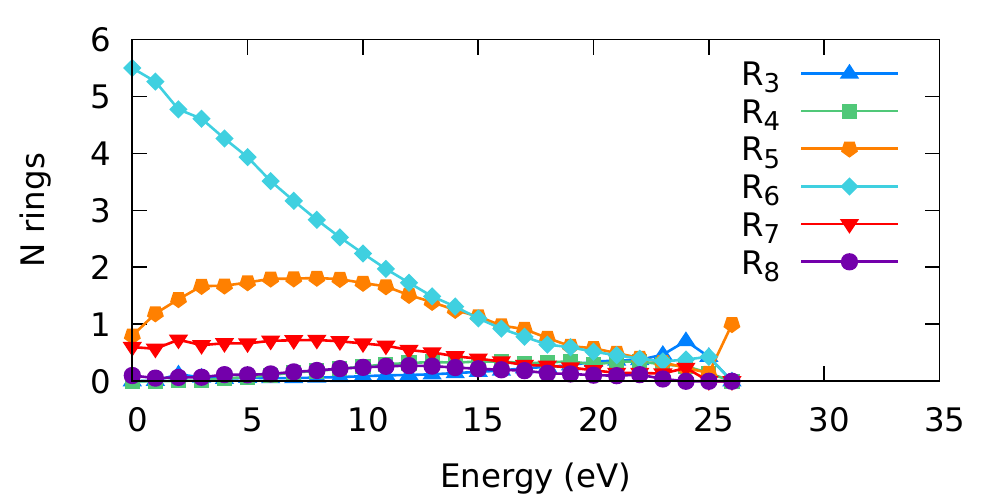}
         \caption{\cbnhm{24}{6}}
         \label{fig:ring_ener_c24h6}
     \end{subfigure}
     \begin{subfigure}[b]{0.49\textwidth}
         \centering
         \includegraphics[width=\textwidth]{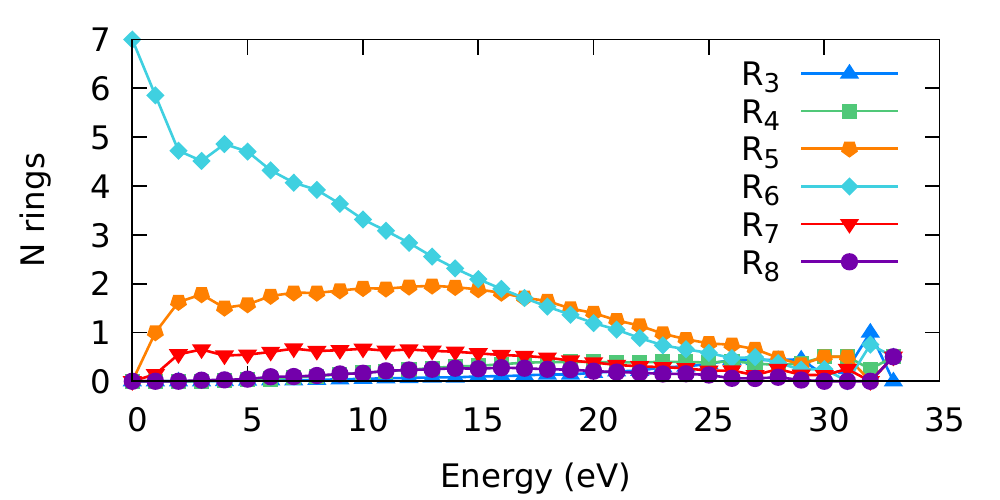}
         \caption{\cbnhm{24}{12}}
         \label{fig:ring_ener_c24h12}
     \end{subfigure}
     \begin{subfigure}[b]{0.49\textwidth}
         \centering
         \includegraphics[width=\textwidth]{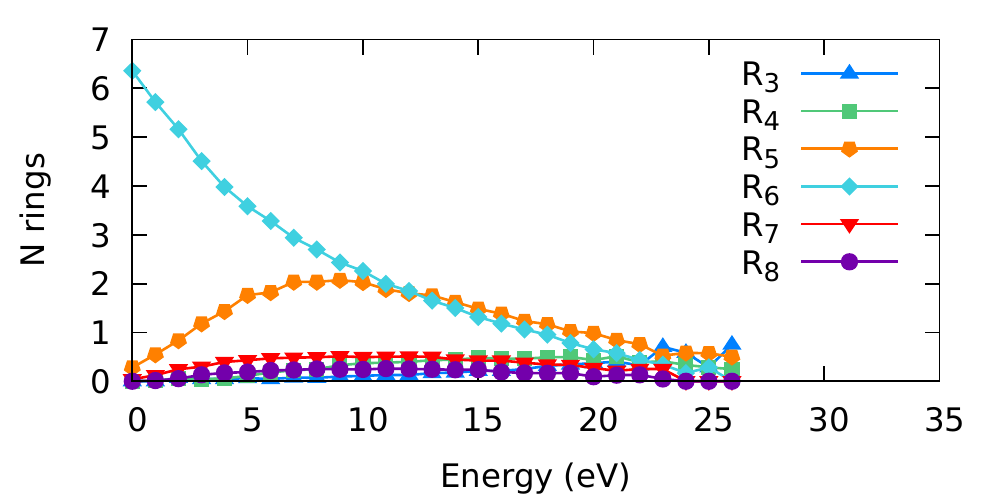}
         \caption{\cbnhm{24}{18}}
         \label{fig:ring_ener_c24h18}
     \end{subfigure}
     \begin{subfigure}[b]{0.49\textwidth}
         \centering
         \includegraphics[width=\textwidth]{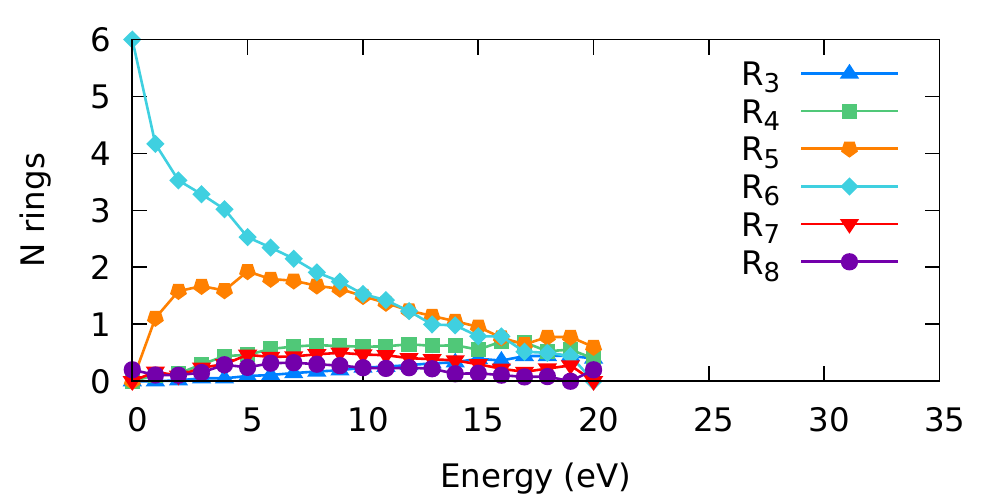}
         \caption{\cbnhm{24}{24}}
         \label{fig:ring_ener_c24h24}
     \end{subfigure}
        \caption{Evolution rings' distributions  as a function of energy for \cbnhm{24}{6} (a), \cbnhm{24}{12} (b), \cbnhm{24}{18} (c) and \cbnhm{24}{24} (d). The data points are obtained averaging among all the structures of each energy window set to 1 eV. Again, the high-energy region is poorly described so points outside the trend might appear. 
 }
        \label{fig:ring_vs_ener}
\end{figure}

We now discuss the distribution of rings' sizes $R_n$ (n = 3-8 C atoms) as a function of energy. As can be seen on Figure~\ref{fig:ring_vs_ener}, the distributions' evolution looks similar for all hydrogenation rates. $R_{5,6}$ are clearly the largest numbers, all the others $R_n$ (n=3,4,7,8) remaining lower than 1. Interestingly, among these minor ring sizes, 7-carbon rings have more occurrences at lower energies for the two most unsaturated hydrogenated carbon clusters ($n_H=6,12$). \\

At lower energies, $R_6$ is clearly major (between 5 and 7) and decreases when energy increases. This indicates that  the existence of 6-carbon rings is one of the main stabilizing factors for C$_m$H$_n$ clusters.  At the lowest energy,  the decrease of $R_6$ is correlated to the increase of $R_5$ and these two rings' sizes disappear at higher energies. For all hydrogenation rates, starting from the minimum energy structures, $R_5$ increases to reach a maximum value of  $2$ and then decreases. Interestingly, $R_5$ remains constant over a wide range of energy ($\sim$ 10 \,eV for the lowest hydrogenation rates, $n_H=6,12$), which is not the case for higher hydrogenation rates ($n_H=18,24$). For the latter, $R_5$  reaches a maximum at $\sim$ 5\,eV, before decreasing similarly to $R_6$.  \\

\subsection{Effects of hydrogenation on ionization energies}

\begin{figure}[htbp!]
    \centering
    \includegraphics[width=0.5\linewidth]{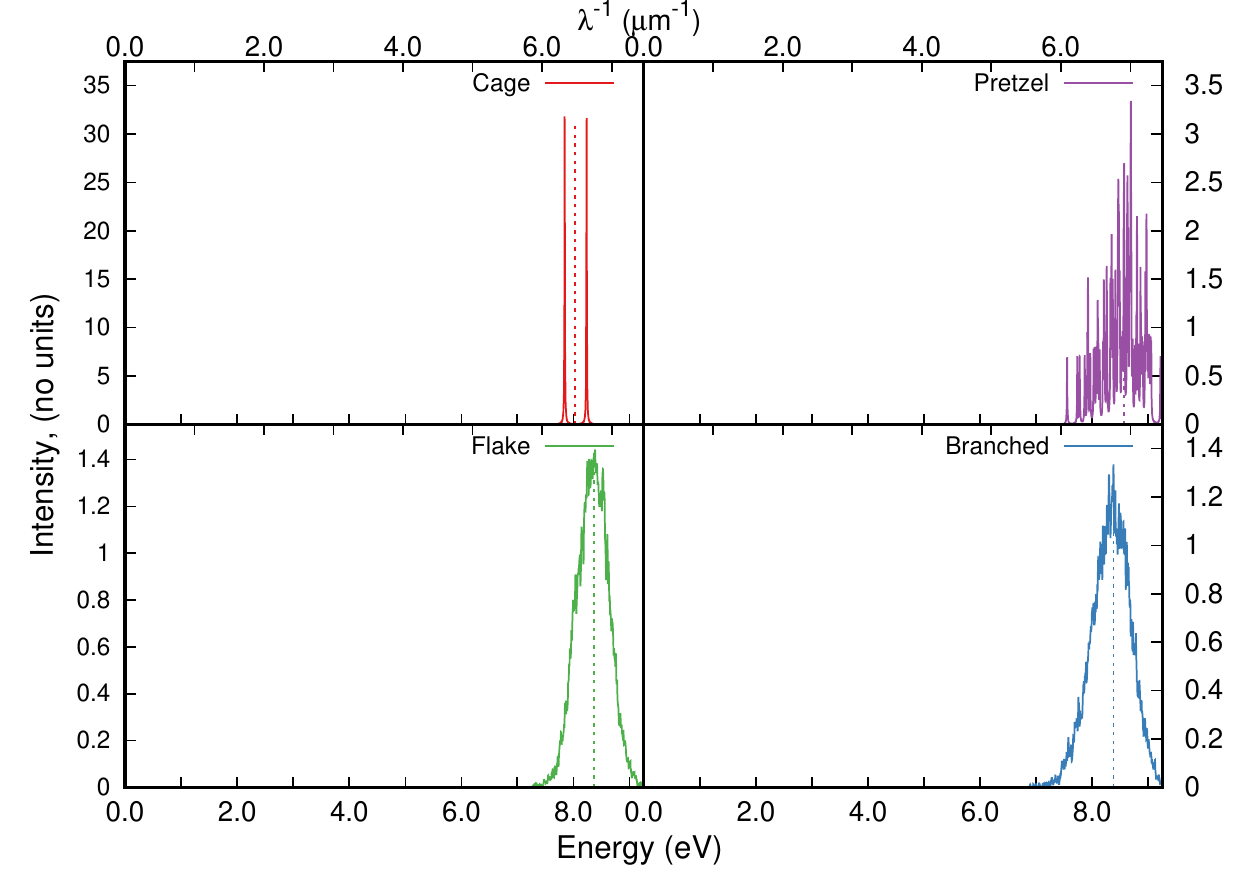} \\
    (a) \\
    \includegraphics[width=0.5\linewidth]{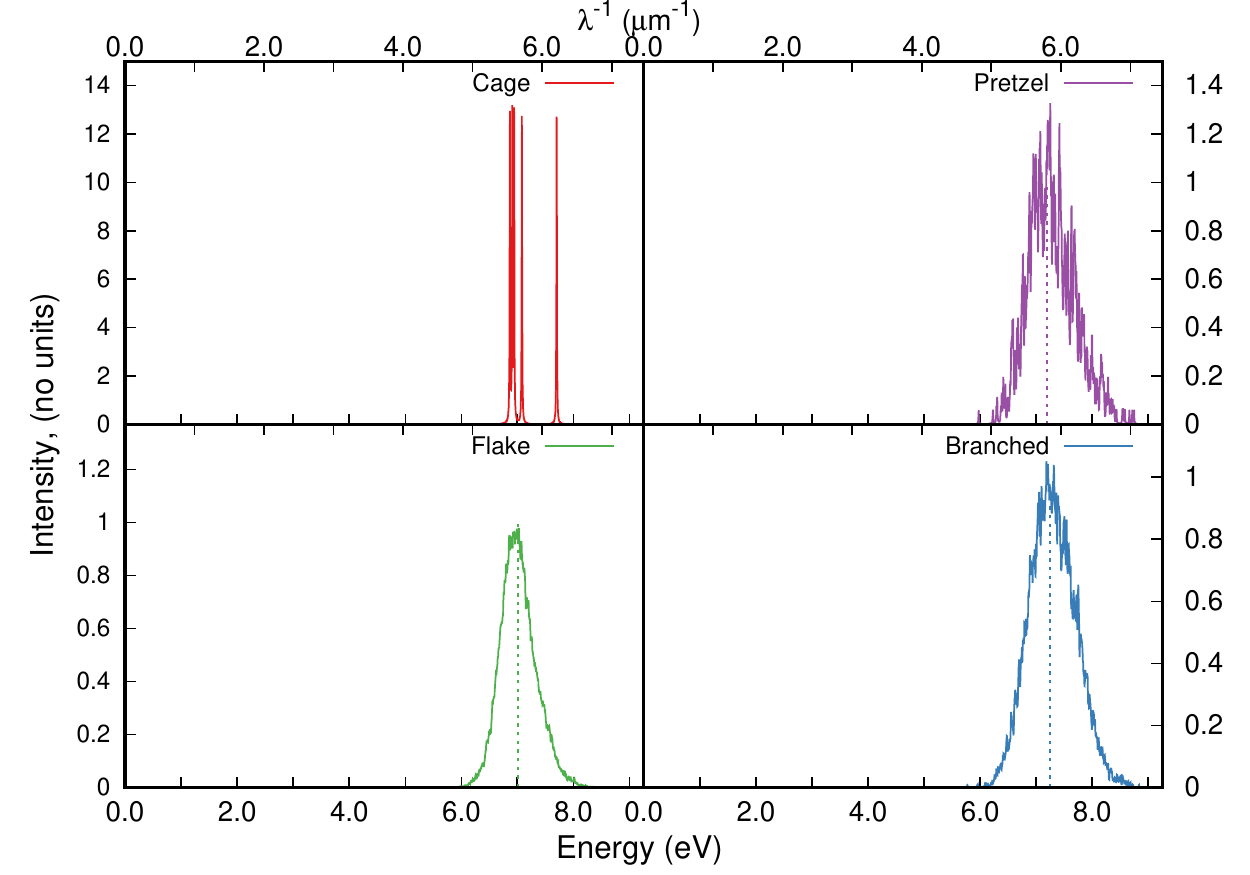} \\
    (b) \\
    \includegraphics[width=0.5\linewidth]{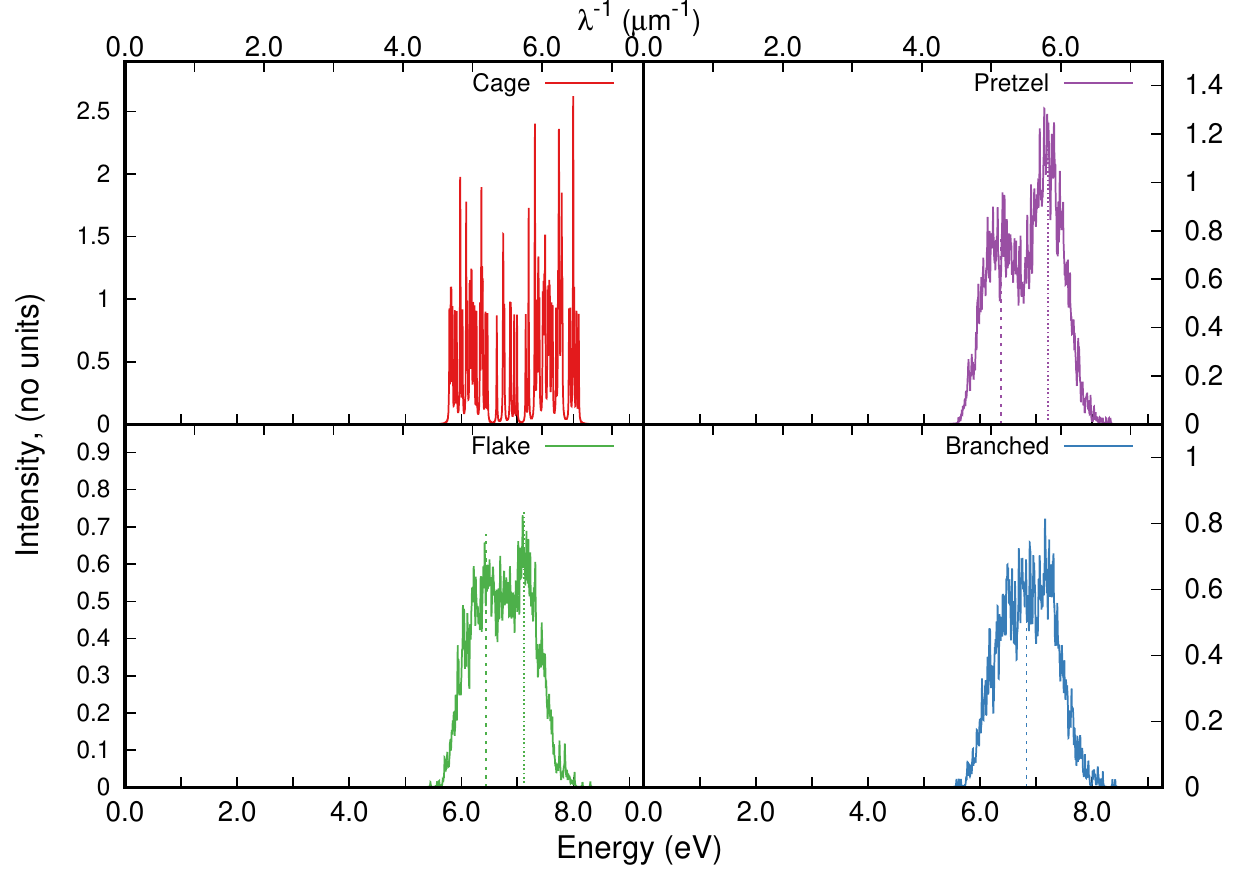} \\
    (c) 
    \caption{Vertical ionization potentials' (VIPs) distributions as a function of energy for the four structural families of \cbn{24} (a) , \cbnhm{24}{12} (b) and  \cbnhm{24}{24} (c). A lorentzian profile of full width at half maximum (FWHM) of 0.01\,eV was applied to the the discrete distribution to improve the appearance.  Normalisation  with respect to the number of isomers per family was also performed.}
    \label{fig:VIP_c24h0-12-24}
\end{figure}

\begin{figure}[htbp!]
    \centering
    \includegraphics[width=0.6\linewidth]{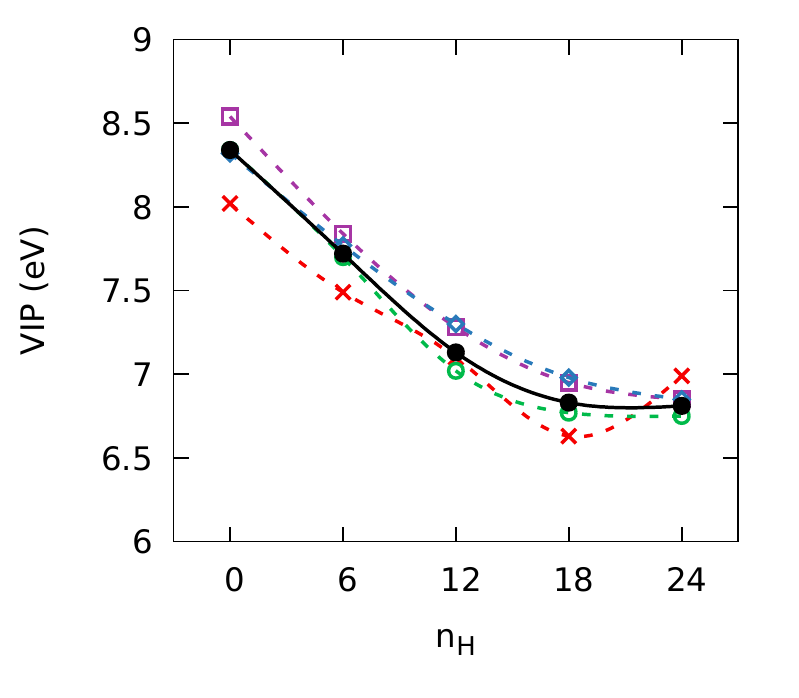} \\
    (a) \\
    \includegraphics[width=0.6\linewidth]{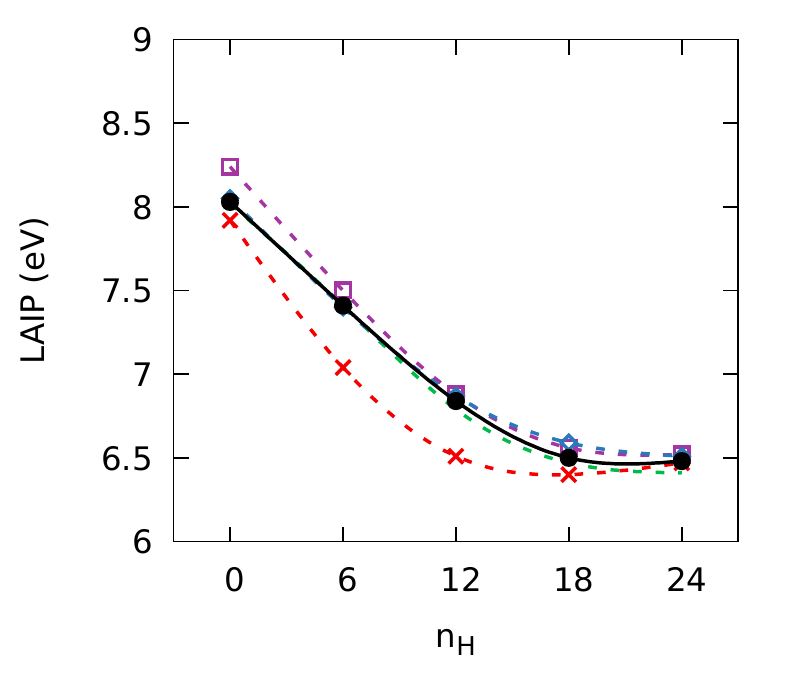} \\
    (b)

\caption{Evolution of the average values of the VIPs( a) and LAIPs(b)  as a function of $n_H$ for the total number of isomers (black full circles) and for each family: cages (red crosses), flakes (green circles), pretzels (purple squares) and branched (blue diamonds). Plain and dashed lines were drawn to guide the eyes.}
\label{fig:evol_IP}
\end{figure}

In this section, we discuss the effect of hydrogenation on the ionization potentials (IPs) of the above discussed families of isomers. Both vertical (VIPs) and "local" adiabatic (LAIPs) ionization potentials were determined for all individual isomers, the former by simply removing one electron of the neutral system and computing the energy difference, the latter by optimizing the cation from the neutral geometry as a starting point, ensuring that it is a minimum, and computing the energy differences. 
 The IPs of isolated PAHs (including coronene) and of PAHs surrounded by water clusters were previously computed at this level of theory (see reference \cite{michoulier2018b}, main text and SI). 
 The VIPs' 
 distribution as a function of energy for all families are reported in Figure~\ref{fig:VIP_c24h0-12-24} for $n_H=0,12, 24$ and in the Supporting Information (Figure~S3) for $n_H=6,18$.   
 The corresponding LAIPs' distribution are reported in the Supporting Information, Figures~S4 and S5 for $n_H$=0,6,12,18 and 24 .
  The average values (per family and over the total number of isomers) of the  VIPs and LAIPs as a function of $n_H$ are reported in Figure~\ref{fig:evol_IP} (a) and (b) respectively. The corresponding data are reported in Table~\ref{tab:AIP_VIP_table_meanvalues}.  The energies corresponding to the distribution maxima for both VIPs and LAIPs are reported in Table~\ref{tab:AIP_VIP_table}. 
\\
  As can be seen in Figure~\ref{fig:VIP_c24h0-12-24} (a), the VIP distributions of \cbn{24} are dominated by an intense band with maxima at similar energies for the flakes and branched families (8.36 and 8.38\,eV, respectively). The VIP distribution for the pretzel family is more noisy due to the smaller number of structures, and the maximum was found at 8.57\,eV. Regarding the cages, the average VIP was found at lower energy (8.02\,eV). Globally, increasing the hydrogenation rate leads to a decrease of the IPs, as detailed below.
As can be seen in Figure~S3, 
 whatever the family, the VIP distribution of \cbnhm{24}{6} is dominated by an intense band, the energy of the maximum depending on the family: it is lower for the flakes (7.62 \,eV) than for pretzels (7.84\,eV) and branched (7.87\,eV). Regarding the cages, one finds discrete transitions due to the few number of structures. Similar as for \cbn{24}, the average value of the VIPs (7.49\,eV) is lower than that of the flakes (7.70\,eV). 
 Increasing the hydrogen ratio results in redshifts of the average IPs (both vertical and adiabatic, see Figure~\ref{fig:evol_IP}) and of the IP's distribution maxima. Interestingly, regarding the average values of the VIPs and of the LAIPs over the total number of isomers for each hydrogenation ratio, the decrease is similar from $n_H=0$ to $n_H=6$ (-0.62\,eV for the VIP) and $n_H=6$ to $n_H=12$ (-0.59\,eV for the VIP), and becomes smaller from  $n_H=12$ to $n_H=18$ (-0.30\,eV for the VIP) up to become negligible from  $n_H=18$ to $n_H=24$ (-0.02\,eV for the VIP). Besides, as can be noticed in Figure~\ref{fig:evol_IP}, the average value over the flakes populations is very similar to the one over all isomers, which was expected as this population is major. For $n_H=24$, the most intense band for the flakes and pretzels families becomes clearly divided into two subbands. The energy of the maxima of the lowest energy subbands follow the trends previously observed for lower hydrogenation rates (redshifts) whereas the other one  is located above 7.0 eV.\\  
 
We can make some attempts to interpret these trends using the evolution of the sp$^n$ carbon fraction described in Section 4.2.3. A decrease in the IP may be due to an increase of the sp$^2$ ratio in conjunction with a decrease of the sp$^1$ ratio. Indeed, referring to model systems, the IP of ethyne (11.40 eV \cite{NIST}) is larger than the IP of ethylene (10.5 eV \cite{NIST}). 
 As previously noticed (Section 4.2.3),  the  sp$^3$ ratio increases with the hydrogenation degree. In this case, it can lead to two opposite trends regarding the IPs. In the first case, the sp$^3$ carbon atoms are under the form of small aliphatic groups such as methyl groups which are electron donor to an aromatic ring (with sp$^2$ electrons). This tends to decrease the IP. For instance the IP of benzene is 9.24 eV \cite{NIST}  whereas that of toluene is 8.83 eV  \cite{NIST}. This trend was confirmed comparing the VIPs of the \cbnhm{24}{6} and \cbnhm{24}{14} flakes isomers of Figure 7, which were determined to be 7.91 and 6.68 eV at the SCC-DFTB level of theory. In the second case, the sp$^3$ carbon atoms are part of carbon cycles. This is expected to lead to an increase of the IP. As an example with model systems, the IP of cyclohexane is 9.88 eV \cite{NIST}), which is 0.62 eV above that of benzene. These trends are confirmed for our systems as the SCC-DFTB  VIPs of the \cbnhm{24}{6} and \cbnhm{24}{14} cages isomers of Figure 7 were found to be 7.55 and 7.96\,eV, respectively.  \\

  Regarding the LAIP spectra (see Supporting Information, Figures~S4 and S5), they are similar to the VIP spectra although : -(i)- they are redshifted, revealing  minor geometrical change upon ionization, with a  value of the shift of the maxima ranging from -0.09\,eV (flakes, \cbnhm{24}{6})  downto -0.24\,eV (branched \cbnhm{24}{18}); -(ii)-  
    additional weaker bands appear at lower energy (below $\sim$ 6\,eV).

\begin{table}[htbp!]
    \centering
    \small
        \begin{tabular*}{\linewidth}{@{\extracolsep{\fill}}ccccccccccc}
     
            \hline\hline
           & \multicolumn{2}{c}{cages} & \multicolumn{2}{c}{flakes} & \multicolumn{2}{c}{pretzels} & \multicolumn{2}{c}{branched}& \multicolumn{2}{c}{total}\\
            \cline{2-3}\cline{4-5}\cline{6-7}\cline{8-9}\cline{10-11}
           
            & A & V & A & V & A &V & A & V &A & V\\ 
            \hline
            \cbn{24} & 7.92 & 8.02 & 8.02 & 8.34 & 8.24 & 8.54 & 8.05 & 8.32 & 8.03&8.34 \\ \hline
            
            \cbnhm{24}{6} & 7.04 & 7.49 & 7.41 & 7.70 & 7.50 & 7.84 & 7.40 & 7.77 &7.41 &7.72 \\ \hline
            
            \cbnhm{24}{12} & 6.51 & 7.10 & 6.79 & 7.02 & 6.88 & 7.28 & 6.87 & 7.30 & 6.84& 7.13 \\ \hline
            
            \cbnhm{24}{18} & 6.40 & 6.63 & 6.47 & 6.77 & 6.56 & 6.95 & 6.59 & 6.98 & 6.50&6.83 \\ \hline

            \cbnhm{24}{24} & 6.47 & 6.99 & 6.41 & 6.75 & 6.52 & 6.85 & 6.51 & 6.85 & 6.48& 6.81 \\ \hline\hline
            
        \end{tabular*}
        \caption{Average value of energies (expressed in eV) found for the LAIPs (A columns) and VIPs (V columns) for all families for all hydrogenation rates. 
        }
        \label{tab:AIP_VIP_table_meanvalues}
\end{table}

\begin{table}[htbp!]
    \centering
    \scriptsize
        \begin{tabular*}{\linewidth}{@{\extracolsep{\fill}}ccccccc}
     
            \hline\hline
           & \multicolumn{2}{c}{flakes} & \multicolumn{2}{c}{pretzels} & \multicolumn{2}{c}{branched}\\
            \cline{2-3}\cline{4-5}\cline{6-7}
           
            & A & V & A &V & A & V\\ 
            \hline
            \cbn{24} &  8.24 (0.74) & 8.36 (0.72) & 8.43 (0.48) & 8.57 (0.48) & 8.28 (0.84) & 8.38 (0.81) \\ 
           & 6.40 (1.31) & ~ & ~ & ~ & 6.27 (1.35) & ~ \\ \hline
            
            \cbnhm{24}{6} &  7.53 (0.84) & 7.62 (0.82) & 7.70 (0.94) & 7.84 (0.61) & 7.68 (1.05) & 7.87 (0.94)\\ 
             & 6.11 (2.41) & ~ & ~ & ~ & 5.83 (2.20) & ~ \\ \hline
            
            \cbnhm{24}{12} & 6.86 (0.79) & 7.01 (0.75) & 7.04 (0.95) & 7.19 (0.87) & 7.11 (0.99) & 7.24 (0.97)\\ \hline
            
            \cbnhm{24}{18} & 
            6.58 (0.89)& 6.73 (0.94) & 6.69 (1.06) & 6.97 (1.08) & 6.69 (1.06) & 6.93 (1.02) \\
            & 4.96 (2.62) &  &  &  &  &  \\ \hline
            
            \cbnhm{24}{24} &  6.24 (0.78) & 6.44 (0.68) & 6.17 (0.67) & 6.37 (0.57) & 6.76 (1.45) & 6.83 (1.30) \\
             & 7.04 (0.66) & 7.12 (0.69) & 7.11 (0.78) & 7.21 (0.76) & ~ &  \\
             & 4.30 (1.09) & ~ & 4.60 (1.12) & ~ & 4.39 (1.29) & ~\\  \hline\hline
            
        \end{tabular*}
        \caption{Energies (expressed in eV, and FWHM in parenthesis) of the maxima found for the LAIPs (A columns) and VIPs (V columns) for all families for all hydrogenation rates. The cage family is not present due to the reduced number of isomers producing discrete bands rather than a continuous distribution.  
        }
        \label{tab:AIP_VIP_table}
\end{table}

In summary, the ionization energies  of the hydrogenated carbonaceous isomers computed in the present work were found lower than those of the most stable well known forms such as coronene (7.3 eV) or  buckminsterfullerene (7.4 eV) at the same level of theory, for $n_H/n_C \geq 1/2$. For a given family, the IP tends to decrease when increasing the hydrogen fraction, probably related to the larger number of isomers with increasing fraction of sp$^2$ carbon atoms with respect to sp$^1$. In the case of $n_H/n_C=1$ where a significant number of sp$^3$ atoms can be found, a second higher energy maximum appears. \\

\section{Conclusion}
In this work, we explored the potential energy surfaces of \cbnhm{24}{n=0,6,12,18,24} up to 20-25\,eV using a genetic algorithm while the electronic structure was determinedat the DFTB0 level of theory, followed by further SCC-DFTB local optimization.  During these simulations, when $n_H/n_C$ increases, enhanced fragmentation was observed with $H_2$ as the major fragment. The structural diversity of the non fragmented structures was analysed using order parameters, that we chose as the number of 5 or 6-carbon rings $Rg(5,6)$ and the asphericity constant $\beta$. 
The most abundant and  lowest energy population correspond to a  flakes population, constituted of isomers of variable shapes (0.3 < $\beta$ $\leq$ 1) with a large number of 5 or 6-carbon rings (3 < $Rg(5,6)$ $\leq$ 8). This population is characterized by a larger number of spherical isomers when $n_H/n_C$ increases.  Simultaneously, the fraction of pretzels structures, more spherical ($\beta$ $\leq$ 0.3) and possessing fewer 5 or 6-carbon ring cycles (0 < $Rg(5,6)$ $\leq$ 8), increases. For all hydrogenation rates, the fraction of cages ($\beta$ $\leq$1.0 and $Rg(5,6)$ $\geq$ 9) remains extremely minor. The branched structures, characterized by the smallest number of 5 or 6-carbon rings ($Rg(5,6)$ $\leq$ 2), is the highest energy population for all  $n_H/n_C$ ratios.  \\
For all hydrogenation rates, we went further into the analysis determining the evolution of carbon hybridization ratio (sp$^n$, n=1-3) as well as rings' sizes distribution  as a function of energy. We showed that sp$^2$ carbon atoms overall dominate, although their predominance decreases when increasing energy due to the increase of $sp^1$ carbon atoms' fraction. Besides, when $n_H/n_C$ increases, the fraction of sp$^3$ carbon atoms increases and remains quite constant over the entire range of considered energies. 
For all hydrogenation ratios, the fraction of 6-carbon rings and, to a lesser extent, of 5-carbon rings, is dominant at low energy but decreases when energy increases.  \\ 
The VIP's distribution of the different populations 
are shown to present a maximum going from 7.87\,eV (\cbnhm{24}{6}, branched family) down to 6.37 eV (\cbnhm{24}{24}, pretzels). The VIPs of branched structures are  systematically higher than that of flakes. When $n_H/n_C$ increases, the VIPs  were found to decrease for all families, and we correlated this trend to the evolution of the geometric and electronic factors studied in the present paper. \\

These results are of astrophysical interest as they should be taken into account in astrophysical models especially regarding the role of carbonaceous species in  the gas ionization. This work will be followed by a report of the IR and UV-visible spectra of the populations analysed in the present work.  \\

\begin{acknowledgement}


The authors gratefully acknowledge financial support by the Agence Nationale de la Recherche (ANR) Grant No. ANR-16-CE29-0025, Spanish MICINN (PID2019-110091GB-I00) and the Severo Ochoa Program for Centres of Excellence in R\&D (SEV-2016-0686). P.P. thanks the Spanish MECD for a FPU grant. E.P. acknowledges the European Union (EU) and Horizon 2020 funding awarded under the Marie Sk\l odowska-Curie action to the EUROPAH consortium, grant number 722346. The authors also thank the computing mesocenter CALMIP (UMS CNRS 3667) for generous allocation of computer resources (p0059 and p17002).

\end{acknowledgement}



\begin{suppinfo}
Definitions of the Hill Wheeler parameters are specified. \\
Figure S1: Distribution of isomers as a function of the sp$^{2}$ ratio for all hydrogenation rates and all families \\
Figure S2: Distribution of isomers as a function of the gyration radius $r_g$ for all hydrogenation rates and all families \\
Figure S3: VIPs' distributions for the four structural families of  \cbnhm{24}{6} and \cbnhm{24}{18} \\
Figure S4: LAIPs' distributions for the four families of structure for \cbn{24}, \cbnhm{24}{6} and \cbnhm{24}{12} \\
Figure S5: LAIPs' distributions for the four families of structure for \cbnhm{24}{18} and \cbnhm{24}{24} \\
All the data computed in the present work are available on the Zenodo general data repository  and associated to the doi: 10.5281/zenodo.4604647  \\

\end{suppinfo}

\providecommand{\latin}[1]{#1}
\providecommand*\mcitethebibliography{\thebibliography}
\csname @ifundefined\endcsname{endmcitethebibliography}
  {\let\endmcitethebibliography\endthebibliography}{}

\end{document}